\documentclass[twocolumn,traditabstract]{aa}

\usepackage[nonamebreak]{natbib}
\usepackage[stable]{footmisc}
\usepackage[fleqn]{amsmath} 
\usepackage{txfonts}
\usepackage{placeins}
\usepackage{natbib}
\bibpunct{(}{)}{;}{a}{}{,}
\usepackage{graphicx}
\usepackage{epstopdf}
\usepackage{ifthen}
\usepackage{mathtools}
\usepackage[breaklinks, colorlinks, citecolor=blue]{hyperref}
\usepackage{url}
\usepackage{comment}
\usepackage[table,usenames,dvipsnames]{xcolor}
\usepackage{xcolor}
\usepackage{gensymb}

\def\setsymbol#1#2{\expandafter\def\csname #1\endcsname{#2}}
\def\getsymbol#1{\csname #1\endcsname}

\newbox\tablebox    \newdimen\tablewidth
\def\leaderfil{\leaders\hbox to 5pt{\hss.\hss}\hfil}
\def\tablenote#1 #2\par{\begingroup \parindent=0.8em
    \abovedisplayshortskip=0pt\belowdisplayshortskip=0pt
    \noindent
    $$\hss\vbox{\hsize\tablewidth \hangindent=\parindent \hangafter=1 \noindent
    \hbox to \parindent{$^#1$\hss}\strut#2\strut\par}\hss$$
    \endgroup}

\def\L2{\ifmmode L_2\else $L_2$\fi}

\def\DeltaT{\ifmmode \Delta T\else $\Delta T$\fi}
\def\deltat{\ifmmode \Delta t\else $\Delta t$\fi}
\def\fknee{\ifmmode f_{\rm knee}\else $f_{\rm knee}$\fi}
\def\Fmax{\ifmmode F_{\rm max}\else $F_{\rm max}$\fi}
\def\solar{\ifmmode{\rm M}_{\mathord\odot}\else${\rm M}_{\mathord\odot}$\fi}
\def\Msolar{\ifmmode{\rm M}_{\mathord\odot}\else${\rm M}_{\mathord\odot}$\fi}
\def\Lsolar{\ifmmode{\rm L}_{\mathord\odot}\else${\rm L}_{\mathord\odot}$\fi}
\def\inv{\ifmmode^{-1}\else$^{-1}$\fi}
\def\mo{\ifmmode^{-1}\else$^{-1}$\fi}
\def\sup#1{\ifmmode ^{\rm #1}\else $^{\rm #1}$\fi}
\def\expo#1{\ifmmode \times 10^{#1}\else $\times 10^{#1}$\fi}
\def\,{\thinspace}
\def\lsim{\mathrel{\raise .4ex\hbox{\rlap{$<$}\lower 1.2ex\hbox{$\sim$}}}}
\def\gsim{\mathrel{\raise .4ex\hbox{\rlap{$>$}\lower 1.2ex\hbox{$\sim$}}}}

\def\simprop{\mathrel{\raise .4ex\hbox{\rlap{$\propto$}\lower 1.2ex\hbox{$\sim$}}}}
\def\deg{\ifmmode^\circ\else$^\circ$\fi}
\def\pdeg{\ifmmode $\setbox0=\hbox{$^{\circ}$}\rlap{\hskip.11\wd0 .}$^{\circ}
          \else \setbox0=\hbox{$^{\circ}$}\rlap{\hskip.11\wd0 .}$^{\circ}$\fi}
\def\arcs{\ifmmode {^{\scriptstyle\prime\prime}}
          \else $^{\scriptstyle\prime\prime}$\fi}
\def\arcm{\ifmmode {^{\scriptstyle\prime}}
          \else $^{\scriptstyle\prime}$\fi}
\newdimen\sa  \newdimen\sb
\def\parcs{\sa=.07em \sb=.03em
     \ifmmode \hbox{\rlap{.}}^{\scriptstyle\prime\kern -\sb\prime}\hbox{\kern -\sa}
     \else \rlap{.}$^{\scriptstyle\prime\kern -\sb\prime}$\kern -\sa\fi}
\def\parcm{\sa=.08em \sb=.03em
     \ifmmode \hbox{\rlap{.}\kern\sa}^{\scriptstyle\prime}\hbox{\kern-\sb}
     \else \rlap{.}\kern\sa$^{\scriptstyle\prime}$\kern-\sb\fi}
\def\ra[#1 #2 #3.#4]{#1\sup{h}#2\sup{m}#3\sup{s}\llap.#4}
\def\dec[#1 #2 #3.#4]{#1\deg#2\arcm#3\arcs\llap.#4}
\def\deco[#1 #2 #3]{#1\deg#2\arcm#3\arcs}
\def\rra[#1 #2]{#1\sup{h}#2\sup{m}}

\def\dots{\relax\ifmmode \ldots\else $\ldots$\fi}
\def\WHzsr{\ifmmode $W\,Hz\mo\,sr\mo$\else W\,Hz\mo\,sr\mo\fi}
\def\mHz{\ifmmode $\,mHz$\else \,mHz\fi}
\def\GHz{\ifmmode $\,GHz$\else \,GHz\fi}
\def\mKs{\ifmmode $\,mK\,s$^{1/2}\else \,mK\,s$^{1/2}$\fi}
\def\muKs{\ifmmode \,\mu$K\,s$^{1/2}\else \,$\mu$K\,s$^{1/2}$\fi}
\def\muKRJs{\ifmmode \,\mu$K$_{\rm RJ}$\,s$^{1/2}\else \,$\mu$K$_{\rm RJ}$\,s$^{1/2}$\fi}
\def\muKHz{\ifmmode \,\mu$K\,Hz$^{-1/2}\else \,$\mu$K\,Hz$^{-1/2}$\fi}
\def\MJysr{\ifmmode \,$MJy\,sr\mo$\else \,MJy\,sr\mo\fi}
\def\MJysrmK{\ifmmode \,$MJy\,sr\mo$\,mK$_{\rm CMB}\mo\else \,MJy\,sr\mo\,mK$_{\rm CMB}\mo$\fi}
\def\microns{\ifmmode \,\mu$m$\else \,$\mu$m\fi}

\def\muK{\ifmmode \,\mu$K$\else \,$\mu$\hbox{K}\fi}
\def\microK{\ifmmode \,\mu$K$\else \,$\mu$\hbox{K}\fi}
\def\muW{\ifmmode \,\mu$W$\else \,$\mu$\hbox{W}\fi}
\def\kms{\ifmmode $\,km\,s$^{-1}\else \,km\,s$^{-1}$\fi}
\def\kmsMpc{\ifmmode $\,\kms\,Mpc\mo$\else \,\kms\,Mpc\mo\fi}
\providecommand{\sorthelp}[1]{}

\def\aap{{A\&A}}

\def\apjs{{ApJSupp.}}

\usepackage{color}
\usepackage{xcolor}

\def\be{\begin{equation}}
\def\ee{\end{equation}}
\def\ba{\begin{eqnarray}}
\def\ea{\end{eqnarray}}

\begin{document}

\title{\vglue -10mm SRoll3: A neural network approach to reduce large-scale systematic effects in the Planck High Frequency Instrument maps
}

\author{G. Wuchterl
	\inst{1}
	\and
	C. Ptolemy\inst{2}\fnmsep\thanks{Just to show the usage
		of the elements in the author field}
}

\institute{Institute for Astronomy (IfA), University of Vienna,
	T\"urkenschanzstrasse 17, A-1180 Vienna\\
	\email{wuchterl@amok.ast.univie.ac.at}
	\and
	University of Alexandria, Department of Geography, ...\\
	\email{c.ptolemy@hipparch.uheaven.space}
	\thanks{The university of heaven temporarily does not
		accept e-mails}
}

\date{Received September 15, 1996; accepted March 16, 1997}

\author{\small
M. Lopez-Radcenco
\inst{1}
\fnmsep
\thanks{Corresponding author: M. ~Lopez-Radcenco,
	\hfill
	\break 
	manuel.lopezradcenco@ias.u-psud.fr}
\and
J.-M. Delouis
\inst{2}
\and
L. Vibert
\inst{1}
}

\institute{\small
 Université Paris-Saclay, CNRS, Institut d'Astrophysique Spatiale, 91405, Orsay, France. \goodbreak \and
Laboratoire d'Océanographie Physique et Spatiale, CNRS, 29238 Plouzan\'{e}, France \goodbreak}

\date{\vglue -1.5mm \today \vglue -5mm}

\abstract{\vglue -3mm 
	In the present work, we propose a neural network based data inversion approach to reduce structured contamination sources, with a particular focus on the mapmaking for Planck High Frequency Instrument data and the removal of large-scale systematic effects within the produced sky maps. The removal of contamination sources is rendered possible by the structured nature of these sources, which is characterized by local spatiotemporal interactions producing couplings between different spatiotemporal scales. We focus on exploring neural networks as a means of exploiting these couplings to learn optimal low-dimensional representations, optimized with respect to the contamination source removal and mapmaking objectives, to achieve robust and effective data inversion. We develop multiple variants of the proposed approach, and consider the inclusion of physics informed constraints and transfer learning techniques. Additionally, we focus on exploiting data augmentation techniques to integrate expert knowledge into an otherwise unsupervised network training approach. We validate the proposed method on Planck High Frequency Instrument 545 GHz Far Side Lobe simulation data, considering ideal and non-ideal cases involving partial, gap-filled and inconsistent datasets, and demonstrate the potential of the neural network based dimensionality reduction to accurately model and remove large-scale systematic effects. We also present an application to real Planck High Frequency Instrument 857 GHz data, which illustrates the relevance of the proposed method to accurately model and capture structured contamination sources, with reported gains of up to one order of magnitude in terms of contamination removal performance. Importantly, the methods developed in this work are to be integrated in a new version of the SRoll algorithm (SRoll3), and we describe here SRoll3 857 GHz detector maps that will be released to the community.

	} 
 

\authorrunning{Lopez-Radcenco et al.}

\titlerunning{SRoll 3: A neural network approach to reduce large-scale systematic effects in the Planck-HFI maps}

\maketitle


\section{Introduction}
\label{sec:intro}
\subsection{Context and motivation}
In the last few decades, scientific instruments have been producing ever increasing quantities of data. Moreover, as remote sensing and instrumentation technology develops, the processing complexity of the produced datasets increases dramatically. The ambitious objectives of several scientific projects are characterized by the reconstruction of the information present in these datasets, which is often mixed with instrumental effects and foreground signals (physical components of the data that mask or blur part of the signal of interest). The scientific community is confronted, in a wide variety of contexts, with the need to extract, from measurements, physical responses adapted to the different models considered, while at the same time ensuring an effective separation between these responses and instrumental effects and/or foreground signals. This separation is rendered possible by the structured nature (in a stochastic sense) of the instrumental effects and/or foreground signals, which, from a mathematical point of view, is characterized by local spatiotemporal interactions producing couplings between different spatiotemporal scales, as opposed to Gaussian signals where no correlation exists between observations produced at different spatiotemporal locations. The structured nature of such signals allows them to be accurately represented using a reduced number of degrees of freedom, which we aim to exploit for their removal from data in order to separate them from the signal of interest, usually less structured and/or Gaussian in nature. 
Given that the aforementioned problem exists in multiple scientific contexts, 
developing efficient dimensionality reduction methods to accurately extract relevant information from data appears as a key issue for the scientific community. 
%
It is therefore essential to identify representations involving a reduced number of degrees of freedom to achieve robust and effective data inversion, while providing enhanced capabilities to accurately describe the complexity of the processes and variabilities at play. In this regard, different strategies can be envisaged, with recent advances relying most notably on the exploitation of operators learned on data presenting some similarities with the problem of interest (e.g. Transfer Learning, as explained below). 
Alternatively, recent works explore the use of generic signal decomposition operators (e.g. Scattering Transform \citep{bruna2015}). Efforts of this type have already yielded interesting results, for example, on the expected statistical description of galactic dust emissions \citep{allys2019}.\\
In the present work, we specifically consider a case study involving the processing of Planck High Frequency Instrument (Planck-HFI) data, with a particular interest on the separation and removal of the systematic effects and foregrounds. 
In this context, we aim here at exploiting machine learning and artificial intelligence approaches to minimize the number of degrees of freedom of the systematic effects to be reconstructed and separated, whereas previous works rely on exploiting spectral and bispectral representations \citep{prunet2001,dore2001,natoli2001,maino2002,degasperis2005,keihanen2005,poutanen2006,armitage-caplan2009,keihanen2010,planck2013-p03f,delouis2019p}, which lack the ability to properly capture spatiotemporal scale interactions, to tackle this issue. Indeed, systematic effects and foregrounds are usually represented using a large number of parameters, whereas more appropriate low-dimensional representations could be learned directly from data. Particularly, in the present work, we focus on exploring neural networks as a means of learning, from data, optimal low-dimensional representations that allow for an effective separation of the structured instrumental effects from the signal of interest, simultaneously with the data inversion. The algorithmic originality of this work lies in the integration of analysis methods issued from machine learning and artificial intelligence to extract the signals of interest from data by minimizing the degrees of freedom of the processes to reconstruct within a classic minimization framework (e.g. a least squares approach). As such, the objective of the proposed methods is to 
find the best low-dimensional description of the data while ensuring an optimal separation of the signal of interest from any instrumental effects and/or foreground signals. We illustrate the relevance of our approach on a case study involving the contamination source removal and mapmaking, i.e, the inversion of raw satellite measurements to produce a physically consistent spatial map, of Planck-HFI data. We consider both Far Side Lobe pickup (FSL) (an unwanted signal due to the non-ideal response of the satellite's antenna) simulations from the 545 GHz Planck-HFI channel and real observations from the 857 GHz Planck-HFI channel. This case study was chosen based on the fact that the FSL pickup is a large-scale systematic effect that currently remains difficult to model and remove, given that the complexity of the Planck optical system forces current FSL estimations to rely on simplified physical and mathematical models. In particular Planck-HFI 545 GHz and 857 GHz channels present 
a weak CMB signature and its sources of contamination are dominated by the FSL pickup, which makes them ideal candidates for the considered case study. Importantly, this work builds on previously developed methods for the separation and removal of structured contamination sources, and particularly on the SRoll2 algorithm \citep{delouis2019p}, used for the production of the 2018 release of the Planck-HFI sky maps. As such, the methods developed in this work are to be integrated in a new version of the SRoll algorithm (SRoll3), and we describe here SRoll3 857 GHz detector maps that will be released to the community. Finally, whereas the application presented provides strong evidence of the relevance of the proposed approach for the processing of large-scale systematic effects, the proposed methodology provides a generic framework for addressing similar, yet complex, data inversion issues involving the separation and removal of structured noise, foregrounds and systematic effects from data in many other scientific domains. 
%
%
\subsection{Related work}
\subsubsection{Convolutional neural networks}
We focus on Convolutional Neural Networks (CNNs), which can be used for extracting relevant information by finding low-dimensional representations of data. To this end, CNNs exploit the existence of invariances within the considered datasets \citep{lecun1998}. 
Broadly speaking, a CNN relies on a cascade of multiple layers, or neurons, to incrementally build an increasingly complex model relating the CNN's inputs and outputs. Each layer consists of a convolution operator with a kernel, whose values are known as weights, followed by the addition of a set of biases. Typically, a non-linear activation function, usually a Regularized Linear Unit (ReLU), i.e., $\text{ReLU}(x)=\max(0,x)$, 
is introduced to allow for non-linear behaviour. Dimensionality reduction/expansion is then achieved by a pooling operator, typically a local averaging or a local maxima. By feeding the output of one layer as input to another layer, multiple layers are then stacked in cascade to build a larger CNN model. Network training then consists on learning, from a training dataset, the network weights and biases that minimize a specific cost function, adequately chosen given the problem of interest. Recent advances have yielded powerful algorithms capable of training large networks from massive datasets efficiently. However, neural network based models are not always invertible, in the sense that part of the (invariant) information fed to the network is lost. This implies that it is not possible to reconstruct 
an input exclusively from the output of a CNN designed to produce a low-dimensional representation of the data. 
Nonetheless, it is indeed possible to synthesize an input that would return a given output when fed to the considered network \citep{mordvintsev2015}. This synthesized input is statistically similar to the original input that produced the output considered (in a sense relating to the neural network architecture and its training). Such results, however, cannot be used to accurately reconstruct the input data, which is why autoencoder networks \citep{bourlard1988,hinton2006} were developed. Autoencoder networks are specifically designed and trained to keep enough information to be able to accurately reconstruct the input data from a low-dimensional representation. Nonetheless, adapting neural network approaches to our application of interest, which closely relates to the problem of source separation in signal processing \citep{choi2005blind}, is not trivial, and would require the imposition of additional constraints on the network weights and biases. Unfortunately, considering additional constraints on the network parameters used for input data reconstruction, which are determined during network training, is not straightforward for autoencoders (or even for most neural networks). 
In order to consider additional constraints, it would be necessary to explicitly rewrite the inversion used by the autoencoder to learn the low-dimensional representation and include any desired constraints within such an inversion scheme. Moreover, autoencoder networks are often based on convolutional approaches that cannot effectively handle partial observations and incomplete data. In this regard, rather than using autoencoder networks, we exploit input-training \citep{tan1995,hassoun1997} to train a deconvolutional decoder network directly from data.
%
\subsubsection{Input-training}
In the present work, both the decoder network parameters as well as the optimal low-dimensional representation of the considered dataset (which constitutes the input of the decoder network) are learned simultaneously during the data inversion, without any preliminary network training phase. The idea of a joint optimization of network parameters and inputs, known as input-training, was first introduced in \citep{tan1995} and later revisited in \citep{hassoun1997} in the context of autoencoder training. Input-training, which closely relates to non-linear principal component analysis \citep{baldi1989,kramer1991,scholkopf1998,scholz2002,scholz2005}, was subsequently exploited for multiple applications, including error and fault detection and diagnosis \citep{reddy1996,reddy1998,jia1998,bohme1999,erguo2002,bouakkaz2012}, chemical process control, monitoring and modeling \citep{bohme1999,liu2004,geng2005,zhu2006}, biogeochemical modeling \citep{somnath2002,schryver2006}, shape representation \citep{park2019} and matrix completion \citep{fan2018}, among others. Recently, this idea was applied in \citep{bojanowski2018} to train generative adversarial networks \citep{goodfellow2014,denton2015,radford2015} without an adversarial training protocol.\\
The choice of an input-training deconvolutional decoder network is further motivated by known limitations of classic CNN-based methods. Indeed, even though CNNs have been extensively used for inverse problems \citep{mccann2017}, most CNN-based approaches learn the optimal solution (in a probabilistic sense) of the considered problem from a very large training dataset that not only needs to accurately represent the complexity of the problem of interest, but that may also not take into account any known and well-understood or well-modeled parts of the underlying processes. As such, CNN-based methods are most effective for the analysis processes where the solution of the inverse problem can be adequately characterized by exploiting a large ensemble of training data. Such approaches usually aim at exploiting a sufficiently large dataset allowing for the development of a complex model capable of generalization to similar observations outside the training dataset. 
In the context of the present study, however, we focus on cases where the signal to be reconstructed is badly known or modeled, and where a limited amount of training data is available. In this respect, we rather aim at exploiting all the available information to produce the most appropriate low-dimensional representation of the available dataset. 
The objective of the decoder network learning stage is then to identify an optimal low-dimensional subspace where both the signal of interest as well as the instrumental effects can be represented accurately, so that the inverse problem can be formulated as a constrained optimization on the projection of these signals onto the learned subspace. The idea is to produce synthesized data from a set of inputs defining a low-dimensional representation of the signals of interest and then compare the synthesized data with real observations. In this regard, the decoder network parameters and the low-dimensional representation are optimized simultaneously, so that the difference between the synthesized dataset and the available observations is minimal.\\
Importantly, this approach is robust to partial observations and incomplete datasets. 
This property is particularly relevant for remote sensing data, which is often derived from satellite or airborne partial surface measurements.
\subsubsection{Transfer Learning}
The basic idea behind transfer learning lies in exploiting knowledge gained by applying machine learning techniques to a specific problem to tackle a different but related problem. Formally, a learning task can be defined by a domain (or dataset) and a learning objective, usually defined by a cost function to be minimized. In transfer learning, knowledge gained from a source learning task is used to improve performance in a different target learning task. This implies that either the domain or the objective of these two distinct tasks are different \citep{pan2009survey}. One may consider, for example, training a galaxy classification algorithm on galaxies from a given survey and then applying the gained knowledge to either classify another set of galaxies from a different survey (different learning domain) \citep{tang2019} or to classify a set of galaxies pertaining to a different classification (different learning objective). It is important to underline that, to be considered as transfer learning, the source and target tasks must be different in either their learning domain and/or their learning objective.\\
The idea of leveraging general knowledge learned from a specific task to improve a similar task is closely related to the concept of generalization. Indeed, using a specific task to extract information that is useful for a secondary task involves identifying specific information that pertains to more general, shared aspects of both tasks. In traditional machine learning approaches, generalization is achieved by building a training dataset that accurately represents a majority of possible cases well enough to generalize to previously unseen observations. In transfer learning, generalization is achieved by means of a more subtle approach that relies on discriminating information specific to the task at hand from general information pertaining globally to both tasks. This may be particularly interesting for the processing of Planck-HFI data, where certain systematic effects are similar between detectors. While this prevents them from being removed by classic averaging-based methods (as they would be accumulated in the mean result used as the final product), it also allows for a very efficient modeling and transfer of shared characteristics between detectors.\\
Transfer learning usually involves training a network to solve the source learning task, and then retraining the last layers of the network on the target learning task. The main idea behind this approach lies in fact that, since the two learning tasks are related, the first layers of the network will involve more general learning pertaining to a more global aspect of the task (like recognizing edges or gradients in image classification), while the final layers exploit this knowledge to build upon it and learn more complex rules.
\subsection{Contributions}
The integration of the decoder network training alongside with the data inversion constitutes the most important original contribution of our approach, as it fundamentally differs from standard dimensionality reduction approaches \citep{kramer1991,demers1993,roweis2000,tenenbaum2000,saul2003,aharon2006,hinton2006,lee2007,van2009,bengio2013}, which are typically used as independent pre-processing steps and produce low-dimensional representations that may not always be completely adapted to the data inversion considered. As such, this dimensionality reduction helps better handle the lack of explicit information on certain instrumental effects and/or foreground signals to effectively separate them from the signal of interest. Finally, particular attention must be paid to the size of the low-dimensional representation, which will directly influence the final number of parameters to be estimated, as a high number of degrees of freedom could adversely affect the identifiability and numerical feasibility of the problem, which can lead to noisy, inaccurate or incorrect solutions. To tackle such an issue one may, for example, consider adding statistical or physically-motivated constraints to the loss function minimized during the data inversion. Here, we illustrate the importance of such dimensionality reduction by considering applications to both synthetic and real Planck-HFI data. In particular, we achieve considerable gains, of up to one order of magnitude, when considering a single input for the low-dimensional representation of the signals of interest.\\
The rest of the paper is organized as follows. In Sect. \ref{sec:method}, we formally introduce the data inversion problem we are interested in, as well as the proposed input-training deconvolutional decoder neural network based formulation, and an alternative two-dimensional formulation of the decoder neural network architecture. Section \ref{sec:applications} introduces applications to both synthetic and real Planck-HFI datasets, provides a comparison to state-of-the-art mapmaking methods and an exploration of the potential of the proposed framework to synthesize and remove FSL pickup. Additionally, it also illustrates how integrating transfer learning techniques into the proposed framework could improve contamination source removal performance. Results pertaining to these applications are presented in Sect. \ref{sec:results} and further discussed in Sect. \ref{sec:discussion}. Finally, we present our concluding remarks and future work perspectives in Sect. \ref{sec:conclusion}.
\section{Method}
\label{sec:method}
\subsection{Problem formulation}
Following standard mapmaking formulations, we cast our data inversion problem as a linear inversion: 
\begin{equation}
 {m}_{t} = {A}_{tp} {s}_{p}+ {c}_{tp}+ {\epsilon}_{t},
\end{equation}
where:
\begin{itemize}
	\item ${m}_t$ is the time-ordered observation data, indexed by a time-dependent index $t$,
	\item ${s}_{p}$ is the spatial signal to recover, indexed by a spatial-dependent index $s$,
	\item ${A}_{tp}$ is a projection matrix relating observations ${m}_t$ to signal ${s}_{p}$, encompassing the observation system's geometry and any raw data pre-processing,
	\item ${c}_{tp}$ is a spatio-temporal dependent signal comprising all structured, non-Gaussian foregrounds and/or systematic effects, 
	\item ${\epsilon}_{t}$ is a time-dependent white noise process modeling instrument measurement uncertainty as well as model uncertainty.
\end{itemize}
The main objective of mapmaking approaches is to recover spatial signal ${s}_p$ from time-ordered observations ${m}_t$, which also involves ensuring an effective separation between ${m}_t$ and foregrounds and systematic effects ${c}_{tp}$, so that there is no cross-contamination in the final produced map.

\subsection{Decoder CNN based inversion method}
\begin{figure}
	\centering
	\includegraphics[width=1\columnwidth]{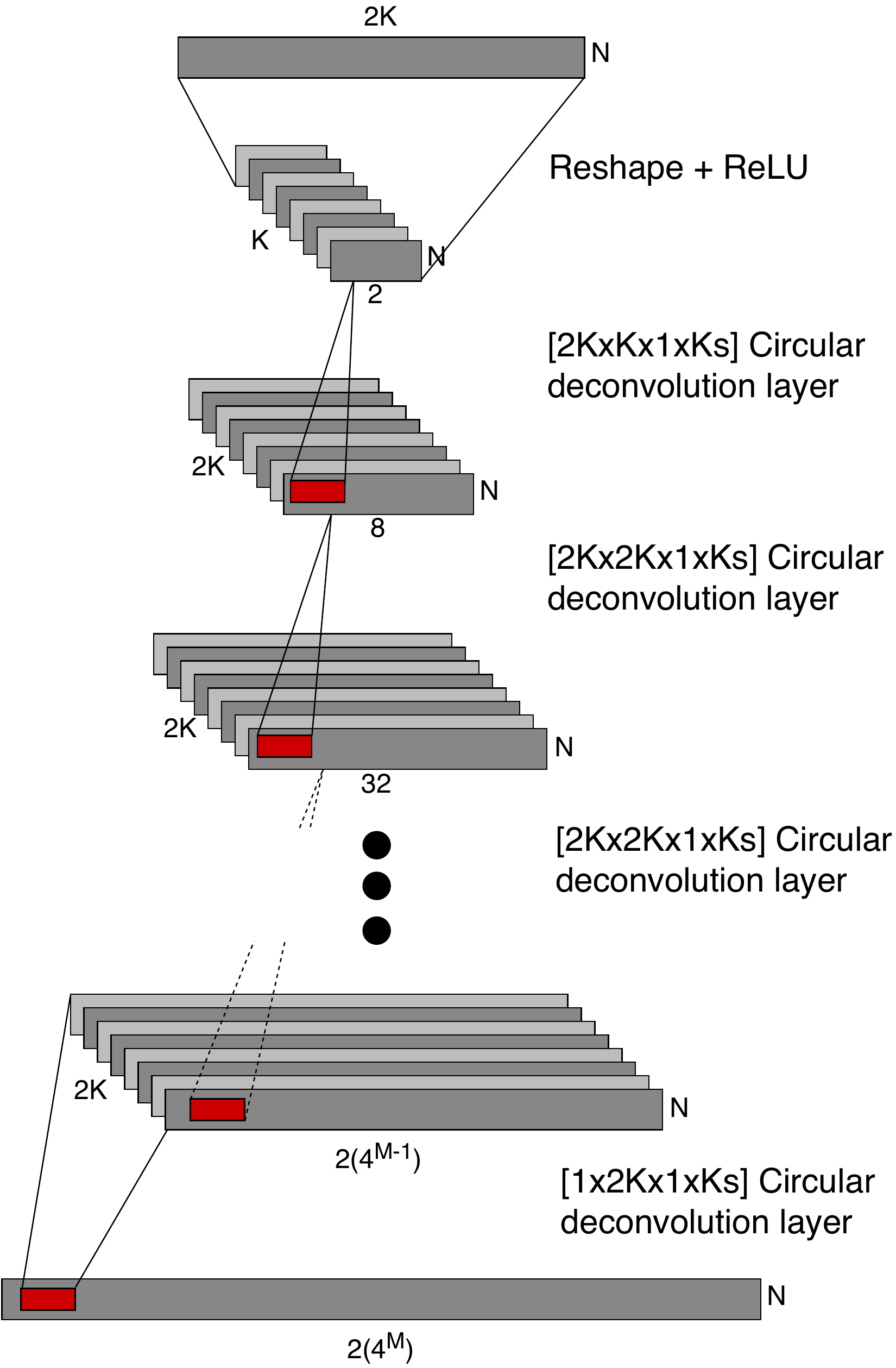}
	\caption{Considered Decoder CNN architecture. 
	}
	\label{fig:decoder_cnn}
\end{figure}

As previously stated, our proposed approach relies on a deconvolutional decoder network to find a low-dimensional representation of foregrounds and systematic effects ${c}_{tp}$, so that it can be effectively separated from spatial signal ${s}_p$. We exploit a custom network training loss function to ensure the effective separation of spatial signal ${s}_p$ from foreground and systematic effects, coupled with an input-training approach to allow for the simultaneous learning of both the network parameters and the low-dimensional representation of ${c}_{tp}$.\\
Specifically, the proposed network architecture takes $N$ low-dimensional feature vectors ${\alpha}_n
,\, n\in\llbracket1,N\rrbracket$ of size $2K$ as input, so that the input data is initially arranged in a 2D tensor of size $[N,2K]$. Input feature vectors are then projected onto a higher-dimensional subspace by means of a deep neural network with multiple deconvolutional layers\footnote{A deconvolutional layer exploits a convolutional kernel to project a low-dimensional input into a higher dimensional subspace by applying an "inverse" convolution (in the sense that the produced output would be projected onto the input by regular convolution with the considered convolutional kernel). Given that a thorough exploration of convolution arithmetic is outside the scope of this work, we refer the reader to \citep{dumoulin2016} for an in-depth analysis of deconvolution in the context of deep neural networks.}. For all feature vectors ${\alpha}_n,\, n\in\llbracket 1,N\rrbracket$, a reshape operation followed by a non-linearity, provided by a ReLU operator, 
converts the input data into $K$ channels of size $n_0=2$, with the result of such operation being a tensor of size $[N,2,K]$. 
A first circular deconvolutional layer 
dilates these $K$ channels into $2K$ channels of sizes $n_1=8$. $M-2$ subsequent convolutional layers 
further dilate these $2K$ channels to sizes $n_2=32,\ldots,n_m=2\cdot4^m,\ldots,n_{M-1}=2\cdot4^{M-1}$, with the corresponding results of such operations being 
tensors of size $[N,8,2K],[N,32,2K],\ldots,[N,2\cdot4^m,2K],\ldots,[N,2\cdot4^{M-1},2K]$, respectively. A final circular deconvolutional layer 
combines the existing $2K$ channels to produce the final output of the decoder network, a 2D tensor ${o}(n,b)$ of size $[N,2\cdot4^M]$. Each of the $N$ lines of this output tensor corresponds to one of the $N$ low-dimensional input feature vectors ${\alpha}_n$.\\
Finally, a piece-wise constant interpolation scheme is used to interpolate the $N$ network outputs of size $2\cdot4^M$ into $N$ outputs corresponding to $N$ higher dimensional output vectors relating to observations ${m}_t$. To this end, for each observation $n$, time-ordered data is binned into $2\cdot4^{M}$ bins, so that all data points corresponding to bin $b$ in observation $n$ are interpolated as ${o}(n,b)$. The binning strategy for this final step is directly dependent on the considered problem and dataset. For Planck-HFI 545 GHz data, this binning is 
detailed in Sect. \ref{sec:applications}. A schematic representation of the network structure is presented in Fig. \ref{fig:decoder_cnn}.\\
As previously explained, neural network based dimensionality reduction is classically performed by exploiting autoencoders, which usually involve deep symmetrical architectures with a bottleneck central layer providing the low-dimensional representation. This is achieved by using observations as both input and output at training, so that the considered network learns the optimal low-dimensional representation space that minimizes reconstruction error. In the proposed approach, however, we rather exploit an input-training scheme to avoid training an encoder network. Input-training is achieved by optimizing the network input, in our case the low-dimensional representation ${\alpha}_n$, alongside with the remaining network parameters. Provided that the considered loss function is differentiable with respect to inputs, classic neural network training approaches can be used to backpropagate gradients through the input layer and optimize the inputs themselves.
\subsubsection{Custom loss function}
In our framework, we wish to ensure an efficient separation between the spatial signal ${s}_p$ and the foregrounds and systematic effects ${c}_{tp}$ modeled by the proposed neural network architecture. To this end, we follow classic mapmaking approaches 
and, under the hypothesis that the projected spatial signal ${A}_{tp} {s}_p$ for any given pixel $p$ remains constant in time, we exploit spatial redundancy in observations ${m}_t$, provided by spatial crossings or co-occurrences in observations at different times, to remove signal ${s}_p$ from observations ${m}_t$. For a given pixel $p$, this comes to computing the mean observation ${M}_p$ and subtracting it from any and all observations ${m}_t$ corresponding to pixel $p$:
\begin{align}
	 {M}_p&=\frac{1}{H(p)}\sum\limits_{\substack{t\\p(t)=p}} {m}_t=\frac{1}{H(p)}\sum\limits_{\substack{t\\p(t)=p}}\left( {A}_{tp} {s}_{p}+ {c}_{tp}+ {\epsilon}_{t}\right) \nonumber \\
	&= {A}_{tp} {s}_{p}+\frac{1}{H(p)}\sum\limits_{\substack{t\\p(t)=p}} {c}_{tp}+ {\epsilon}_{t},
\end{align}
\begin{equation}
\hat{ {m}}_t= {m}_t- {M}_p= {m}_t-\frac{1}{H(p)}\sum\limits_{\substack{t\\p(t)=p}} {m}_t= {c}_{tp}-\frac{1}{H(p)}\sum\limits_{\substack{t\\p(t)=p}} {c}_{tp},
\label{eq:mapmaking}
\end{equation}
where $p(t)$ designates the pixel corresponding to observation ${m}_t$ at time $t$, and $H(p)$, known as the hit-count, is the total number of observations at pixel $p$.\\
In the proposed decoder network based approach, observations ${m}_t$ are used for training by considering the output of the decoder network to provide a parametrization of the foregrounds and systematic effects:
\begin{equation}
	 {c}_{tp}=f\left( {\alpha}_n\right),
	\label{eq:dcnn}
\end{equation}
so that the network, including its inputs ${\alpha}_n$, can be trained to minimize reconstruction error with respect to signal free observations $\hat{ {m}_t}$. The appropriate training loss function can then be directly derived from Eqs. (\ref{eq:mapmaking}) and (\ref{eq:dcnn}):
\begin{align}
	\mathcal{L}&=\sum\limits_{p}\sum\limits_{\substack{t\\p(t)=p}}\left\vert\left\vert\left( {m}_t-\frac{1}{H(p)}\sum\limits_{\substack{t\\p(t)=p}} {m}_t\right)- \left( {c}_{tp}-\frac{1}{H(p)}\sum\limits_{\substack{t\\p(t)=p}} {c}_{tp}\right)\right\vert\right\vert^2_2 \nonumber\\
	&=\sum\limits_{p}\sum\limits_{\substack{t\\p(t)=p}}\left\vert\left\vert\left( {m}_t- {M}_p\right)- \left( {c}_{tp}-\frac{1}{H(p)}\sum\limits_{\substack{t\\p(t)=p}} {c}_{tp}\right)\right\vert\right\vert^2_2 \nonumber\\
	&=\sum\limits_{p}\sum\limits_{\substack{t\\p(t)=p}}\left\vert\left\vert\left( {m}_t- {M}_p\right)- \left(f\left( {\alpha}_n\right)-\frac{1}{H(p)}\sum\limits_{\substack{t\\p(t)=p}}f\left( {\alpha}_n\right)\right)\right\vert\right\vert^2_2.
	\label{eq:loss}
\end{align}
From Eqs. (\ref{eq:mapmaking}), (\ref{eq:dcnn}), and (\ref{eq:loss}), it can be concluded that the time invariance hypothesis of projected spatial signal ${A}_{tp} {s}_p$ ensures that all traces of signal ${s}_p$ can be adequately removed from observations ${m}_t$ during the data inversion. Even though this hypothesis may not always be formally respected depending on the considered dataset and application, it still remains a valid approximation for a large number of applications, provided that the appropriate spatio-temporal scales and sampling frequencies for observations are chosen.\\
Following recent trends in machine learning \citep{raissi2017a,raissi2017b,karpatne2017,lusch2018,nabian2018,raissi2018a,raissi2018b,raissi2018c,yang2018,erichson2019,lutter2019,roscher2019,seo2019,yang2019}, we design a custom loss function including a standard reconstruction error term (as in most machine learning applications) coupled with physically-derived terms introducing expert knowledge relating to the application and dataset considered (see Sect. \ref{sec:applications} for detailed examples).
\subsubsection{2D Decoder CNN alternative formulation}
\label{sec:2d}
Besides the previously introduced Decoder CNN architecture, we propose an alternative two-dimensional (2D) formulation of the original Decoder CNN. The novel 2D formulation amounts to modifying the network so that the intermediate convolutional layers involve two-dimensional convolutional kernels. In this regard, this alternative formulation relies on a two-dimensional binning of observations ${m}_t$ for training. In particular, we exploit here a fully connected layer to allow us to considerably reduce the dimension of the low-dimensional representation of the signals of interest, at the expense of increasing the number of weights and biases to be learned during training. Contrary to a convolutional layer, which involves a convolution where each value of the produced multidimensional output depends only on a local subset of a multidimensional input (due to the convolution operation), a fully connected layer produces the output by means of a linear combination of all values in the input. In a fully connected layer, the weights and biases to be learned are those of the linear combination that produces the output. This implies that trainable inputs, i.e., the low-dimensional representation, ${\alpha}_n,\, n\in\llbracket 1,N \rrbracket$ of size $K$ should now be arranged into a 2D tensor of size $[N,K]$, which will be converted by a fully connected layer into $K$ channels of size $[2,2]$, with the result of such operation being a tensor of size $[2,2,K]$. A first convolutional layer 
further expands this tensor into $2K$ channels, producing an output of size $[8,8,2K]$. In a similar fashion to the original Decoder CNN, $M-2$ subsequent circular deconvolutional layers 
dilate these $2K$ channels along the first two dimensions into sizes $n_1=8,n_2=32,\ldots,n_m=2,\cdot4^m,\ldots,n_{M-1}=2\cdot4^{M-1}$, with the corresponding results of such operations being 
tensors of size $[8,8,K],[32,32,K],\ldots,[2\cdot4^m,2\cdot4^m,K],\ldots,[2\cdot4^{M-1},2\cdot4^{M-1},K]$, respectively. A final circular deconvolutional layer 
combines the existing $2K$ channels to produce a tensor of size $[2\cdot4^M,2\cdot4^M]$.\\
Finally, a piece-wise constant interpolation scheme is used to interpolate the network outputs of size $[2\cdot4^M,2\cdot4^M]$ into outputs corresponding to observations ${m}_t$. To this end, time ordered data is binned two-dimensionally into $(2\cdot4^{M})\times(2\cdot4^{M})$ bins. 
The binning strategy for this final step is directly dependent on the considered problem and dataset. 
\subsection{Map constraint}
\label{sec:map_constraint}
Given that the proposed approach exploits spatial redundancy in the observations by minimizing loss function \ref{eq:loss}, which is computed on observation co-occurrences only, no strong constraint is imposed on the large-scale signature of the network output. In this regard, the network output may, in some cases, resort to adding a large-scale signal that remains close to zero around the ecliptical poles where most signals crossings occur, in order to further minimize the loss function. Since few crossings exist in between the ecliptical poles, this large-scale signal will not be adequately constrained by observations and will rarely produce a physically sound reconstruction. To prevent such behaviour, the following additional constraint on the final correction map, given by $\sum\limits_{\substack{t\\p(t)=p}}f\left( {\alpha}_n\right)$ is considered:

\begin{equation}
	\mathcal{L}_{map}=\sum\limits_{p} \left\vert\left\vert {M}_p- \frac{1}{H(p)}\sum\limits_{\substack{t\\p(t)=p}}f\left( {\alpha}_n\right)\right\vert\right\vert^2_2.
	\label{eq:constraint_map}
\end{equation}

Such constraint will penalize solutions where the final correction map diverges from the input map, thus avoiding the inclusion of a strong large-scale signature on the network correction.\\
The compromise between the original loss function (\ref{eq:loss}) and the additional map constraint (\ref{eq:constraint_map}) is controlled by means of a user-set weight $W_{map}$, so that the final modified loss function is given by:
\begin{align}
	\mathcal{L}_{total}=&\mathcal{L}+W_{map}\mathcal{L}_{map}\nonumber\\
	=&\sum\limits_{p}\sum\limits_{\substack{t\\p(t)=p}}\left\vert\left\vert\left( {m}_t- {M}_p\right)- \left(f\left( {\alpha}_n\right)-\frac{1}{H(p)}\sum\limits_{\substack{t\\p(t)=p}}f\left( {\alpha}_n\right)\right)\right\vert\right\vert^2_2\nonumber\\
	&+W_{map}\sum\limits_{p} \left\vert\left\vert {M}_p- \frac{1}{H(p)}\sum\limits_{\substack{t\\p(t)=p}}f\left( {\alpha}_n\right)\right\vert\right\vert^2_2.
	\label{eq:loss_total}
\end{align}
%
%
\subsection{Transfer learning}
\label{subsubsec:tl}
In the context of the processing of Planck-HFI observations, transfer learning techniques appear as particularly interesting given the limited amount of data available. 
This is in perfect agreement with the main motivation behind transfer learning, which aims at leveraging knowledge from previously learned models to tackle new tasks, thus going beyond specific learning task and domains to discover more general knowledge shared among different problems. As illustrated below, transfer learning allows us to do so by using data from different bandwidths and exploit different detectors to complement each other and produce more accurate 
sky maps.\\
To further constraint the proposed decoder network, particularly for identifying and removing contamination sources shared among multiple detectors, we explore classic transfer learning techniques. 
As previously explained, transfer learning relies on learning and storing knowledge from a particular problem or case study and applying such knowledge to solve a similar but different problem or case study.\\
Given the specificities of the proposed Decoder CNN architecture, we train the whole network on a source task and only retrain the low-dimensional representation (i.e., the low-dimensional inputs ${\alpha}_n$) on the target task. Such an approach can be seen as a particular case of feature-representation transfer learning \citep{pan2009survey}, since the knowledge transferred between tasks lies in the way the signals and processes of interest are represented in the low-dimensional subspace of the inputs. Indeed, since we may consider the Decoder CNN as a projection of the observations onto the low-dimensional space of the inputs, transferring the network weights and biases and only retraining the inputs amounts to considering that the projection onto the space of the inputs is shared between the two learning tasks considered. This means that the source learning task will learn a projection, defining a low-dimensional representation, that will then be used as is by the target task.\\
In the context of the proposed application, we exploit transfer learning to better learn structured systematic effects and/or foregrounds by training the proposed Decoder CNN on a dataset accurately depicting these foregrounds and systematic effects. Given that we focus on learning the projection that most accurately captures the structure of the foregrounds and systematic effects, the Decoder CNN is trained on the whole dataset rather than on observation co-occurrences (as is done with cost function (\ref{eq:loss})), which amounts to considering the following training cost function:

\begin{equation}
	\mathcal{L}_{TL}=\sum\limits_{p}\sum\limits_{\substack{t\\p(t)=p}}\left\vert\left\vert {m}_t-{c}_{tp}\right\vert\right\vert^2_2 =\sum\limits_{p}\sum\limits_{\substack{t\\p(t)=p}}\left\vert\left\vert {m}_t- f\left( {\alpha}_n\right)\right\vert\right\vert^2_2.
	\label{eq:loss_tl}
\end{equation}

After training, the resulting Decoder CNN is transferred to a new dataset, which amounts to retraining the Decoder CNN inputs only using the original cost function (Eq. (\ref{eq:loss})), while keeping the previously trained weights and biases.\\
In this context, two distinct cases can be discerned. The first case involves two detectors that measure the sky signal in the same frequency band, while the second case involves two detector measuring the sky signal on different frequency bands. Both cases rely on training the Decoder CNN on a dataset from a specific detector, and then, retraining inputs only on a different dataset pertaining to a different detector. Since the learning tasks for both detectors are different (the considered cost functions are different), such a procedure effectively amounts to transfer learning. This is further reinforced if the second detector dataset differs significantly from the first detector dataset, i.e, if we choose, for example, to train the Decoder CNN on a 545 GHz detector dataset and retrain the inputs using a 857 GHz detector dataset. The simpler case where both detectors measure the sky signal in the same frequency band still amounts to transfer learning, but may involve more accurate knowledge transfer, given the strong similarities between the source and target datasets.
\section{Applications}
\label{sec:applications}
%

\subsection{Planck observation strategy}
\begin{figure}
	\centering
	\includegraphics[width=1\columnwidth]{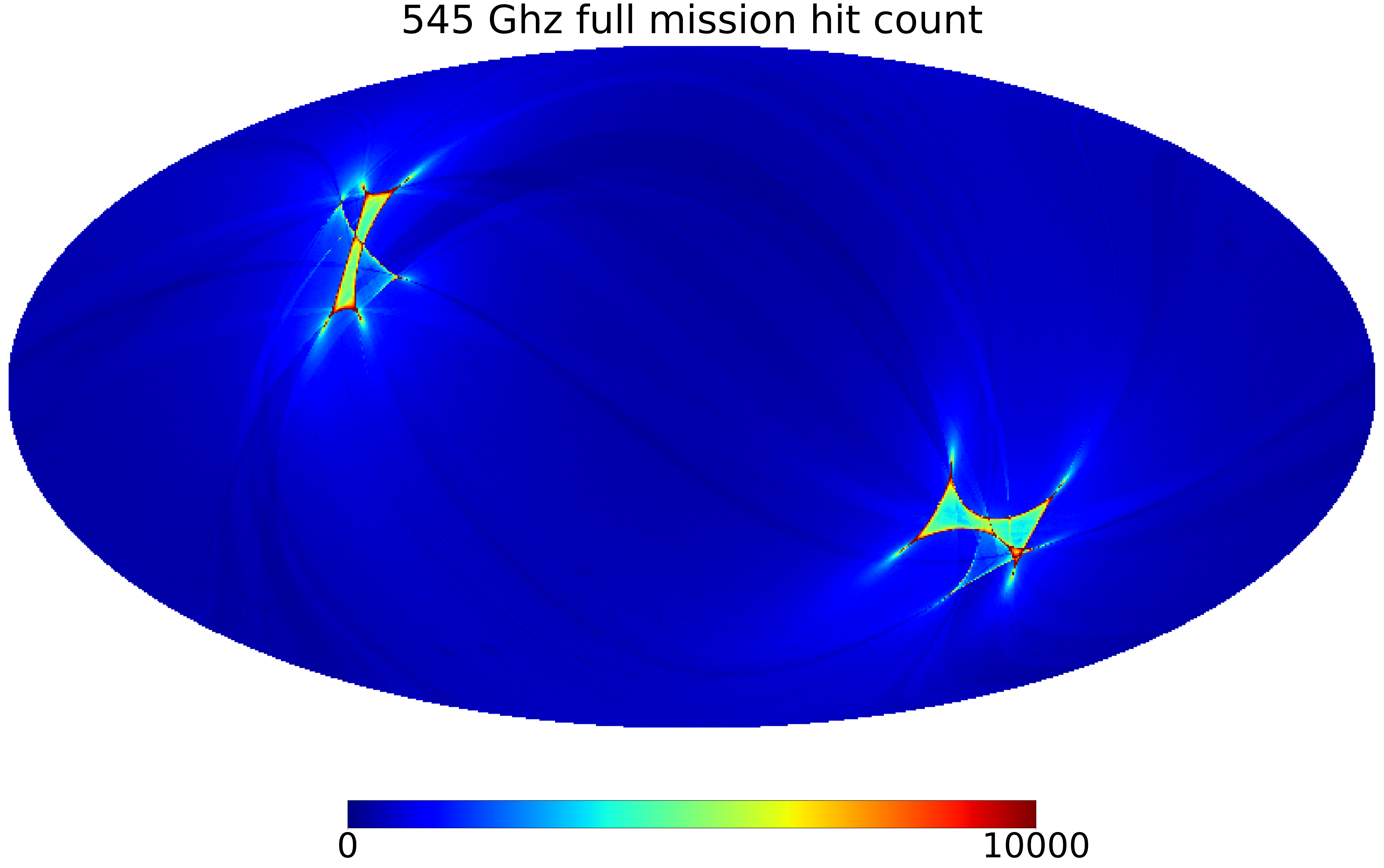}
	\caption{Full-mission observation hit-count map, i.e, the total number of observations at each pixel, for the Planck-HFI 545 GHz channel.}
	\label{fig:hc545}
\end{figure}

The Planck satellite scanning strategy, a clear schema of which can be found in Sect. 1.4 of \cite{planck2016-ES}, is determined by a halo orbit around the Lagrange L2 point. The satellite rotates around an axis nearly perpendicular to the Sun \citep{tauber2010a} and scans the sky in nearly great circles at around 1 rpm, which means that the ecliptic poles are observed considerably more frequently, and in many more directions, than the ecliptic equator. Thus, ecliptic poles concentrate most of the observation crossings and co-occurrences providing the required redundancy to ensure effective separation and removal of systematic effects and foregrounds. 
This can be clearly observed in Fig. \ref{fig:hc545}, where we present Planck-HFI 545 GHz channel hit-count map, i.e, the number of observations at each pixel.\\
The redundancy pattern produced by the Planck-HFI scanning strategy is particularly relevant for our approach, given that the network training takes spatial redundancy into account for the removal of signal $s_p$ to ensure that the CNN is trained to capture and model foregrounds and systematic effects $c_{tp}=f(\alpha_n)$ only. In this regard, the choice of a scanning strategy is a critical point in the design of most remote sensing satellite missions, as it determines a compromise between spatial redundancy (necessary for an accurate removal of spatially redundant sources of contamination) and spatiotemporal sampling resolution (necessary to obtain accurate and reliable measurements of the signal of interest). For most remote sensing satellite mission design, the choice of scanning strategy is usually the product of extensive research based on multiple end-to-end simulations of the observing system.
%
%
\subsection{Planck data pre-processing and compression}
\label{subsec:compression}
Time ordered data from the Planck satellite is sampled in consecutive 1 rpm rotations of the satellite. 
These observations can be naturally organized into discrete packages, with measurements corresponding to each full rotation being grouped together into units called circles. Given the relationship between the rotational velocity of the satellite and its orbital velocity around the Sun, consecutive circles can be grouped together every 60 rotations and averaged to produce a composite measurement, called a ring, under the approximation that the region of the sky observed by 60 consecutive circles ($\sim 1$ hour) remains constant.\\
Given that a ring corresponds to 60 averaged rotations at a constant angular velocity, the sampling frequency of the instrument then determines a uniform sampling of the phase space within each ring. In this way, the phase of a full rotation is discretized into $B$ points, so that each measurement in a ring corresponds unequivocally to a phase bin of amplitude $\frac{2\pi}{B}$.\\
Further compression of the information present in the Planck-HFI 545 GHz and 857 GHz datasets is achieved by considering a HEALPix pixelization \citep{gorski2005} with $N_\text{side}=2048$ and averaging, for each ring, all measurements that fall within the same pixel. As far as phase information is concerned, each new averaged measurement is associated with a composite phase value obtained by averaging the phase of all measurements falling within the considered pixel. In this way, phase information loss due to averaging, and the associated sub-pixel artifacts (i.e., inconsistent pixel values related to the loss of phase information), are minimized. Since the considered compression stage produces results of varying length depending on each ring's orientation with respect to the pixelization grid, zero padding is used to produce a homogeneous dataset by converting all rings to length $l=27664$, which is the length of the largest compressed ring in the dataset.
\subsection{Far Side Lobe pickup large-scale systematic effect}
For the application considered in the present work, we focus specifically on one large angular scale systematic effect, namely the FSL pickup, which consists of radiation pickups far from the Planck telescope line of sight, primarily due to the existence of secondary lobes in the telescope's beam pattern, which creates what is commonly known as ``straylight contamination'' \citep{tauber2010a,planck2014-a04}. Typically, FSL pickup is characterized by a highly structured large angular scale signature, which makes it an ideal candidate to evaluate the proposed method's ability to exploit such structure to project the signals of interest onto a low-dimensional subspace where such structured information is adequately represented with a reduced number of degrees of freedom. In this regard, we focus our analysis on larger spatial scales (below multipole $\ell=100$, i.e., angular scales over $1\degree$), given that FSL pickup is primarily a large-scale systematic effect. Moreover, the dominant contamination source at small scales in Planck-HFI data is the detector noise, which can be modeled as an unstructured, Gaussian signal that cannot be effectively removed by the proposed method, which further motivates our choice to focus on large spatial scales. It should be noted, however, that even though we focus here on the FSL pickup specifically, other structured contamination sources present in intermediate spatial scales not yet dominated by detector noise may also be removed with the proposed methodology, but this is beyond the scope of this work.
\subsection{Planck-HFI 545 GHz dataset}
To illustrate the relevance of the framework introduced in Sect. \ref{sec:method}, we consider data from the Planck-HFI 545 GHz dataset 
of the Planck mission \citep{tauber2010a}. As previously explained, the choice of the Planck-HFI 545 GHz channel is motivated by its 
weak CMB signature, which simplifies both the data processing and the interpretation of obtained results. In particular, we exploit FSL pickup synthetic 545 GHz data to validate our method's ability to learn suitable low-dimensional representations of the FSL pickup under both ideal and non-ideal settings, including cases considering incomplete, gap-filled and inconsistent datasets.

\subsection{Planck-HFI 857 GHz dataset}
Besides Planck 545 GHz data, we also consider Planck 857 GHz data to evaluate how data augmentation techniques 
can be exploited to improve the contamination source removal performance of the Decoder CNN architecture, as explained in Sect. \ref{sec:results_857}. Similarly to the Planck-HFI 545 GHz channel, the Planck-HFI 545 GHz channel presents 
a weak CMB signature, which simplifies both the data processing and the interpretation of obtained results.\\
Importantly, the Planck-HFI 857 GHz channel has the particularity that, given the position of its associated detectors on the focal plane, detector $857_2$ presents very little FSL pickup. This implies that the detector difference maps between different detectors will predominantly depict large-scale systematic effects, with the FSL pickup being the dominant systematic observed \citep{planck2016-l03p}. As such, Planck-HFI 857 GHz data provides an ideal setting to evaluate the ability of the proposed method to capture and remove large-scale systematic effects, and specially the FSL pickup. Moreover, the considered Planck 857 GHz data share many similarities with the previously introduced Planck 545 GHz dataset, including the circle-averaging used to produce rings and the HEALPix pixelization based information compression (presented in Sect. \ref{subsec:compression}). Besides the difference in frequency bands, the main difference lies in the slightly different observation spatial distribution, produced by the differences in location and orientation of the detectors involved.
%
%
%
%
%
%
\subsection{Decoder network training}
Taking Planck-HFI 545 GHz and 857 GHz data specificities into consideration for the proposed decoder network based approach, we chose to train our decoder network on compressed rings directly, so that the final step of the decoder network, considering $M=3$, uses a piece-wise constant interpolation to interpolate $l=27664$ values from the $2(4^3)=128$ larger bins produced as output by the decoder network. Using phase values as the independent interpolation variable, network outputs are thus interpolated to length $l=27664$, and compared to compressed rings during training. Network parameters and inputs are jointly optimized in order to minimize reconstruction error while ensuring an effective 
removal of systematic effects and foregrounds by minimizing the custom loss function (Eq. (\ref{eq:loss})). Moreover, an additional map constraint term, as introduced in Sect. \ref{sec:map_constraint}, is added to the loss function (Eq. (\ref{eq:loss})) to introduce physics informed constraints and leverage domain knowledge on the inversion problem. Finally, transfer learning strategies, as presented in Sect. \ref{subsubsec:tl} are also explored as a means to share and transfer relevant information between datasets.
\section{Results}
%
\label{sec:results}
\subsection{Validation on 545 GHz FSL simulations}
\label{sec:tl_results}
We explore the ideas introduced in Sect. \ref{sec:applications} by training our Decoder CNN on FSL simulation data from a 545 GHz detector (detector $545_1$). The objective of this validation stage is to demonstrate the method's ability to adequately learn a suitable low-dimensional representation for the signals of interest from data. In this regard, the learned representation embeds knowledge, learnt from the available dataset, that facilitates the separation and removal of structured systematic effects and foregrounds (provided that such knowledge effectively exists within the dataset), while also being optimized with respect to the data inversion itself. Moreover, transfer learning techniques are evaluated using phase-shifted 
data from the same detector. In this regard, considering phase-shifted data allows us to simulate either partially similar detector datasets, and/or cases where the FSL pickup is partially or badly modeled. It should be noted that the use phase-shifted data is purely exploited as a means to emulate missing knowledge within the training dataset. In this regard, the considered phase shift is not necessarily a representation of the real physical phenomena occurring within the satellite's optical system, but it is rather a simplified scheme to demonstrate how the proposed inversion method responds to training on incomplete or inconsistent datasets.\\
All considered datasets consist of $747\,093\,984$ observations packed into rings of size 27664, for a total ring count of $N=27006$ rings.\\
We evaluate performance by presenting and comparing results obtained for the following approaches:
\begin{itemize}
	\item A classic destriping \citep{planck2014-a09} of detector $545_1$ FSL simulation data (referred to as CD hereafter),
	\item A direct fit of a FSL template computed from $20\degree$ phase-shifted detector $545_1$ FSL simulation data onto detector $545_1$ FSL simulation data (referred to as TFIT hereafter),
	\item The Decoder CNN in its original 1D version trained and applied directly on detector $545_1$ FSL simulation data (referred to as CNN1D hereafter),
	\item The Decoder CNN in its original 1D version trained and applied directly on detector $545_1$ FSL simulation data and considering an additional weighted map constraint (referred to as CNN1D-$W_{map}$ hereafter),
	\item The Decoder CNN in its original 1D version trained on $20\degree$ phase-shifted detector $545_1$ FSL simulation data and applied to non-shifted detector $545_1$ FSL simulation data by retraining inputs only (referred to as CNN1D-TL hereafter).
\end{itemize}

Subsequently, we also perform additional tests to evaluate:

\begin{itemize}
	\item The performance of the 2D alternative formulation of the Decoder CNN for the original case studies and datasets (referred to as, respectively, CNN2D, CNN2D-$W_{map}$ and CNN2D-TL hereafter),
	\item The performance of the proposed algorithms when applied to a gap-filled dataset generated by subsampling available observations.
\end{itemize}
For comparison and benchmarking purposes, we include, among the methods considered, a classic destriping approach. This result is used as baseline for evaluating the performance of the proposed methodology. For the considered methods, results are evaluated quantitatively by means of final full mission output maps and half mission difference maps, which are presented for visual comparison.\\
Regarding the processing of Planck-HFI data, several ways of splitting the datasets for their analysis are described in \citep{planck2016-l03p}. Here, we use half mission difference maps, which are computed by dividing the whole time ordered data series in two equal halves, processing each half independently and then computing the difference between the obtained maps. As such, half mission difference maps remove all spatially redundant information, allowing for the analysis of the information remaining once structured spatial signals are removed. This implies that half mission difference maps provide relevant information regarding the training of the Decoder CNN, since it is trained using a custom cost function that explicitly removes reduntant spatial information, but they do not provide much information regarding real contamination source removal performance.\\
Moreover, a quantitative performance evaluation is given by means of the power spectra of the presented maps, which are computed using a spherical harmonics decomposition. In this spectral representation, multipole scale number $\ell$ relates to different spatial angular scales. As such, the power spectra depicts how energy is distributed across angular scales, thus providing a multi-scale measurement of the power per surface unit within the analyzed map.
\subsubsection{1D and 2D Decoder CNN}
\begin{figure}
	\centering
	\includegraphics[width=1\columnwidth]{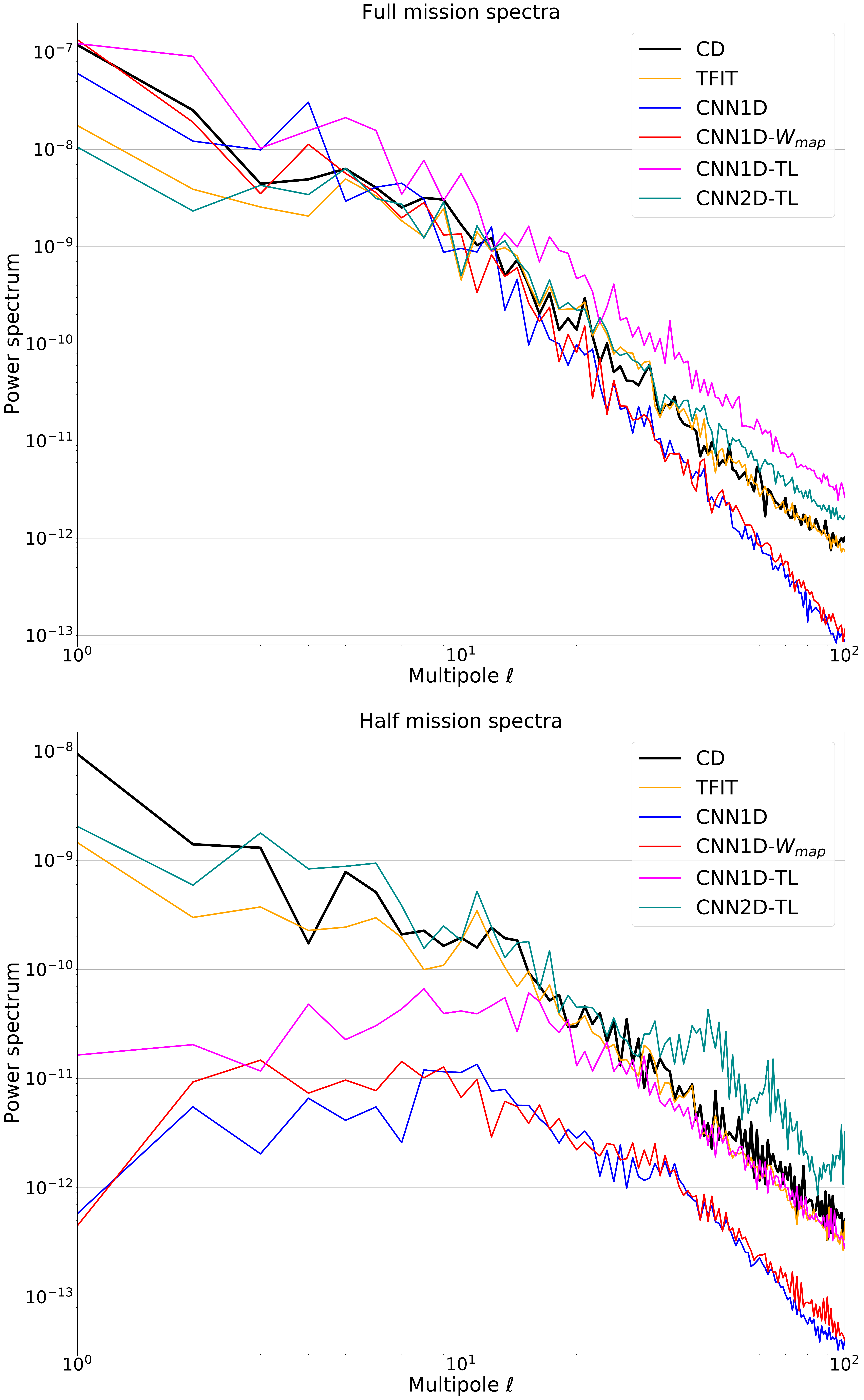}
	\caption{Power spectra (in arbitrary units) of full mission maps (top) and half mission difference maps (bottom) of detector $545_1$ FSL simulations after contamination source removal using 1000 iterations of a classic destriping approach (CD), a direct fit of phase-shifted detector $545_1$ FSL simulation data as a template (TFIT), the original 1D Decoder CNN (CNN1D), the 1D Decoder CNN variants using the additional map constraint (CNN1D-$W_{map}$) and transfer learning by training the Decoder CNN weights and biases on $20\degree$ phase-shifted data from detector $545_1$ FSL simulations (CNN1D-TL), and the 2D variant of the Decoder CNN using transfer learning by training the Decoder CNN weights and biases on $20\degree$ phase-shifted data from detector $545_1$ FSL simulations (CNN2D-TL).}
	\label{fig:fullhalf}
\end{figure}

\begin{figure*}
	\centering
	\includegraphics[width=1\textwidth]{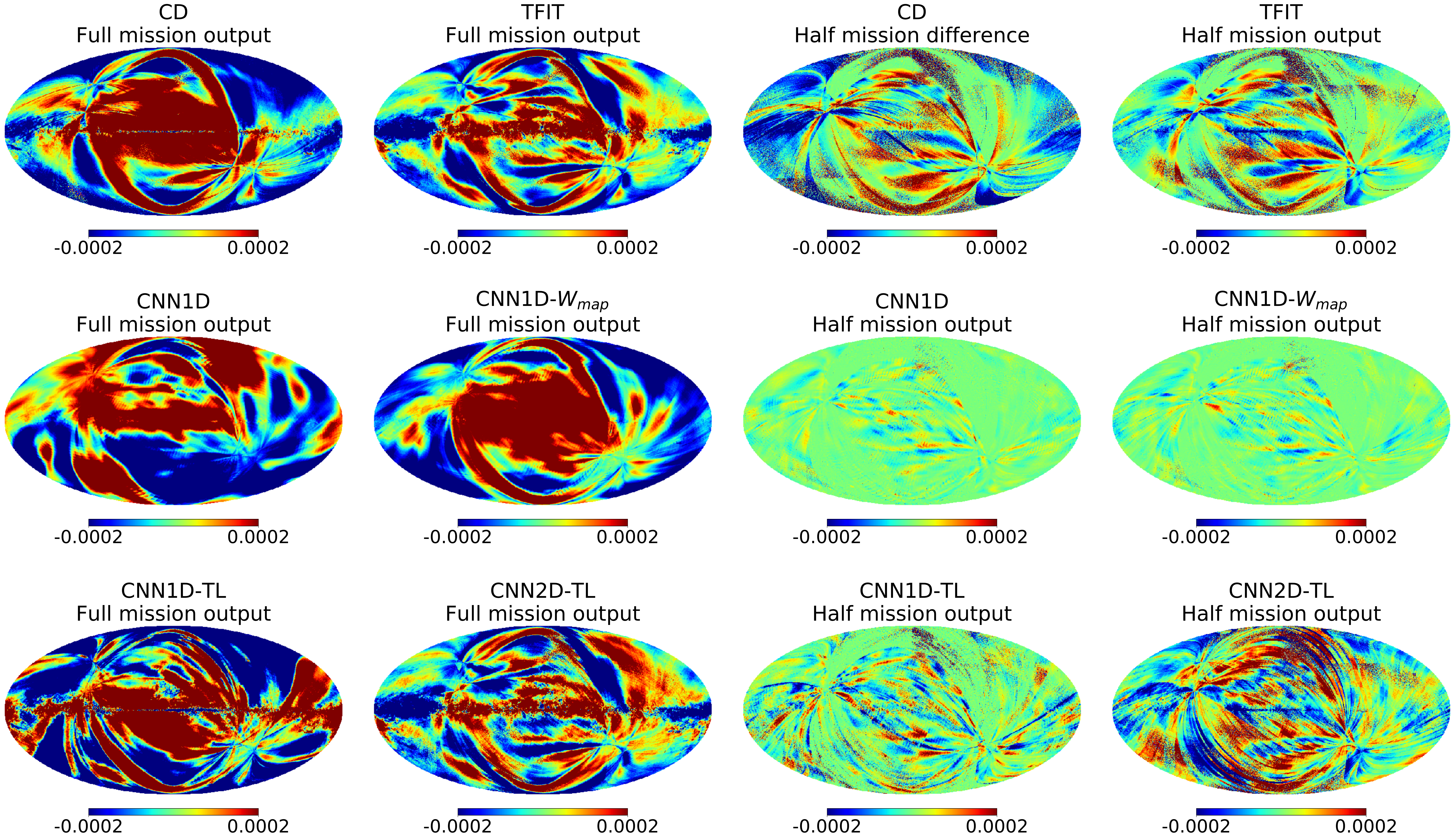}
	\caption{Full mission and half mission difference maps (in arbitrary units) of detector $545_1$ FSL simulations after contamination source removal using 1000 iterations of a classic destriping approach (CD), a direct fit of phase-shifted detector $545_1$ FSL simulation data as a template (TFIT), the original 1D Decoder CNN (CNN1D), the 1D Decoder CNN variants using the additional map constraint (CNN1D-$W_{map}$) and transfer learning by training the Decoder CNN weights and biases on $20\degree$ phase-shifted data from detector $545_1$ FSL simulations (CNN1D-TL), and the 2D variant of the Decoder CNN using transfer learning by training the Decoder CNN weights and biases on $20\degree$ phase-shifted data from detector $545_1$ FSL simulations (CNN2D-TL). Leftmost columns present full mission maps, rightmost columns present half mission difference maps.}
	\label{fig:fullmaps}
\end{figure*}

The first considered case study involves exploiting data from the Planck-HFI 545 GHz channel only. In this context, transfer learning amounts to training the Decoder CNN on phase-shifted data from detector $545_1$, and exploiting this network to process non-shifted data from detector $545_1$ by retraining inputs only. As previously stated, such an approach can be considered as transfer learning despite the similarities between the two datasets, since the source and target learning tasks are different.\\
For the considered case study, the 1D Decoder CNN considers $K=4$ channels and $M=3$, so that the 1D Decoder CNN architecture uses 4 deconvolutional layers to project a total number of $8N$ inputs onto time ordered data binned into $128$ phase bins. For the map constrained version of the Decoder CNN (trained with loss function (\ref{eq:loss_total})), we consider $W_{map}=10^{-2}$, which was chosen empirically as it produced the best results when testing the method's sensitivity to this parameter.\\
We further complement our performance study by analyzing and comparing results obtained by contamination source removal FSL simulation data with the 2D variants of the Decoder CNN introduced in Sect. \ref{sec:2d}. To this end, we exploit a 2D Decoder CNN to process 545 GHz data under identical conditions as those analyzed for the 1D Decoder CNN. 
In this regard, the results for the 2D CNN Decoder were also obtained by considering $K=4$ and $M=3$, so that 
the considered 2D Decoder CNN consists of four 2D deconvolutional layers and will project $16\, (4\times4)$ inputs onto time ordered data vectors binned into $128\times128$ bins in phase and time. For the map-constrained version, user-set weight $W_{map}$ is once again set to $W_{map}=10^{-2}$.\\
Figure \ref{fig:fullhalf} presents, for the different considered approaches, the power spectra of the full mission maps and the half mission difference maps for the different variants considered. For a qualitative analysis of these results, Fig. \ref{fig:fullmaps} presents these full mission and half mission difference maps themselves. Additionally, Figs. \ref{fig:fullhalf} and \ref{fig:fullmaps} also includes maps, and their corresponding power spectra, for the best result obtained when exploiting the 2D formulation of the Decoder CNN, i.e., for CNN2D-TL (exploiting CNN weights and biases learned on $20\degree$ phase-shifted detector $545_1$ FSL simulation data). For the sake of simplicity and readability, the lesser performing variants of the 2D formulation are not included in Figs. \ref{fig:fullhalf} and \ref{fig:fullmaps}. Moreover, since we consider here idealized synthetic simulation data, numerical results have no real physical interpretation, and are thus presented using arbitrary units.\\
Concerning the 1D variants of the Decoder CNN, Fig. \ref{fig:fullhalf}, 
shows that both CNN1D and CNN1D-$W_{map}$ provide a substantial gain for the filtering of smaller scale FSL structures, while not being able to accurately remove the large-scale FSL signature. 
CNN1D, however, obtains the best performance in terms of large-scale contamination source removal (at multipole $\ell=0$). CNN1D-TL (trained on a phase-shifted detector dataset) degrades performance overall, while TFIT provides the best results at larger scales while not being able to capture smaller scale structures. As far as the 2D variant of the Decoder CNN is concerned, it seems that CCN2D-TL considerably improves on contamination source removal performance for larger spatial scales, even outperforming TFIT. For smaller spatial scales, however, the use of CNN2D-TL does not seem to provide any performance gain, with respect to CD, and may even degrade performance for larger $\ell$ values. 
Similarly, from Fig. \ref{fig:fullhalf}, one can observe a considerable gain for all spatial scales in the half mission difference maps when considering CNN1D and CCN1D-$W_{map}$. Globally, CNN1D seems to provide the more substantial gain for most spatial scales. In agreement with these findings, the half mission difference map for CNN1D is less energetic and closer to a Gaussian white noise in space (even though some residual signal can still be observed) than the other analyzed Decoder CNN variants. 
The use of CNN1D-TL, however, does seem to provide some gain for all spatial scales, even though it is marginal when compared to CNN1D and CNN1D-$W_{map}$, specially for smaller spatial scales. TFIT produces an even smaller performance gain, remaining quite close to the performance levels of CD, while CNN2D-TL appears as the worst performing variant, overall, for half mission difference map contamination source removal. Indeed, CNN2D-TL, the best performing 2D variant of the Decoder CNN, does not seem to provide much gain in contamination source removal performance for half mission difference maps, with a performance level slightly worse than TFIT at larger spatial scales, and a clear degradation in contamination source removal performance, with respect to a CD, for smaller spatial scales.
Overall, the half mission difference maps are in strong agreement with this analysis.\\
Given that previous results showed little degradation in terms of half mission difference maps contamination source removal performance, results presented in the following sections focus specifically on full mission contamination source removal performance. Moreover, for the sake of readability, we focus exclusively on a quantitative analysis by means of power spectral plots, and do not include additional map plots.
\subsubsection{Partial observations with large gaps}
\begin{figure}
	\centering
	\includegraphics[width=1\columnwidth]{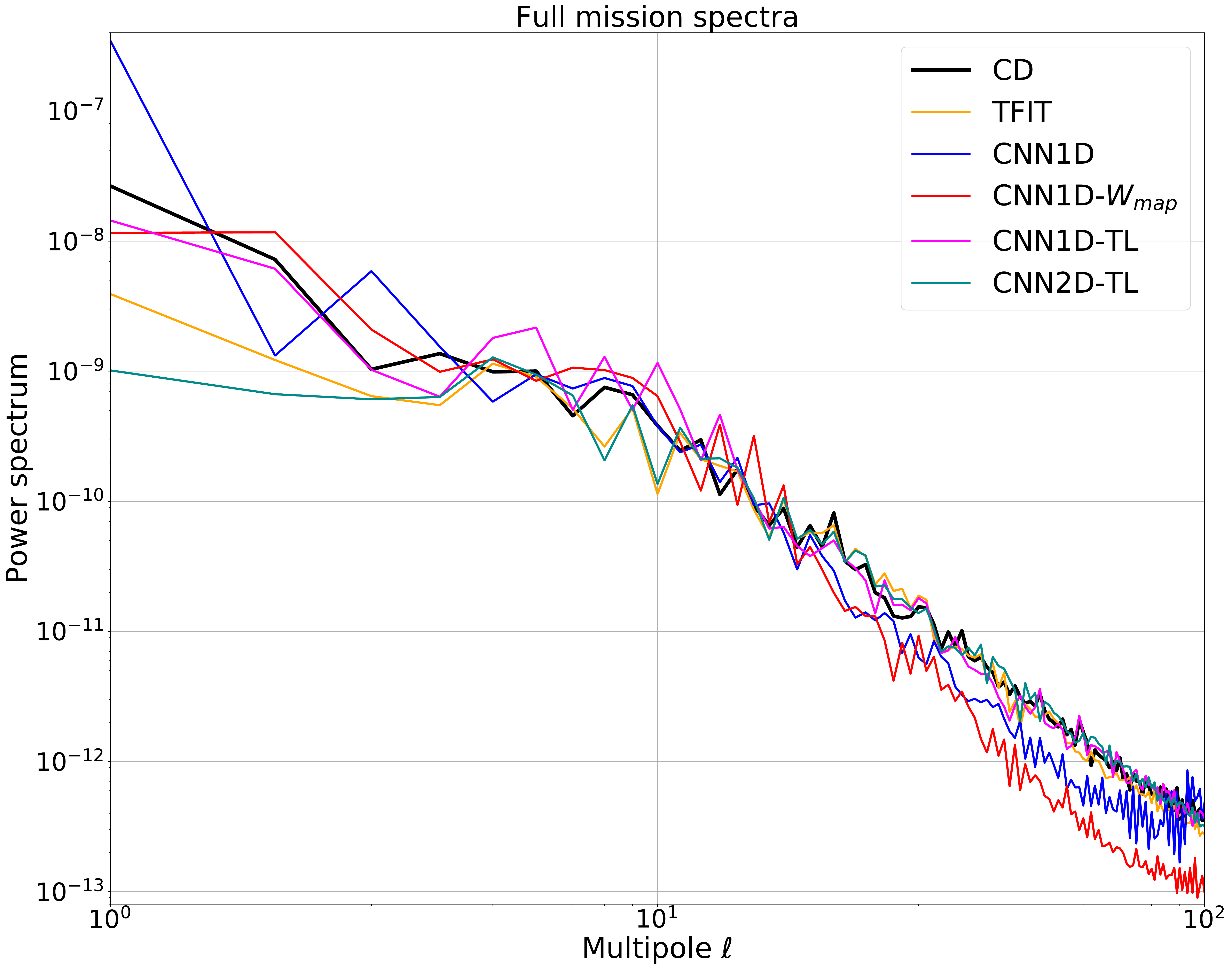}
	\caption{Power spectra (in arbitrary units) of full mission maps of detector $545_1$ FSL simulations considering one every ten rings after contamination source removal using 1000 iterations of a classic destriping approach (CL), a direct fit of phase-shifted detector $545_1$ FSL simulation data as a template (TFIT), the original 1D Decoder CNN (CNN1D), the 1D Decoder CNN variants using the additional map constraint (CNN1D-$W_{map}$) and transfer learning by training the Decoder CNN weights and biases on $20\degree$ phase-shifted data from detector $545_1$ FSL simulations (CNN1D-TL), and the 2D variant of the Decoder CNN using transfer learning by training the Decoder CNN weights and biases on $20\degree$ phase-shifted data from detector $545_1$ FSL simulations (CNN2D-TL).}
	\label{fig:fullhalf_r10}
\end{figure}

To further illustrate the relevance of transfer learning techniques, we now consider the previously introduced Planck-HFI 545 GHz dataset but sub-sample one every ten rings, which amounts to considering a partial dataset involving large gaps. We consider an identical configuration for the considered Decoder CNN as the one used for previously presented results, namely $K=4$ channels and $M=3$, for a total of $128$ phase bins for the 1D Decoder CNN and $128\times128$ bins in phase and time for the 2D Decoder CNN. For the map constrained versions of the Decoder CNN, $W_{map}$ is kept at its original value of $W_{map}=10^{-2}$.\\
We present similar results as those introduced in Sect. \ref{sec:tl_results}. i.e., full mission maps 
power spectra in Fig. \ref{fig:fullhalf_r10}. 
Our initial analysis of the obtained results indicate that, given the large gaps in the considered dataset, the Decoder CNN tends to add a considerable spatial offset to the whole map in order to fill in those gaps. During our tests, this effect was partially limited by the additional map constraint, even though this does suffice to completely remove the offset. From the full mission maps before removing offsets, we observed that both CNN1D and CNN1D-$W_{map}$ were unable to correctly capture and filter the FSL signal. CNN1D-TL, however, considerably improves performance when considering partial datasets involving large gaps, most notably for smaller spatial frequencies.\\
After subtracting the spatial mean, we observe that performance is considerably improved, particularly for CNN1D-TL, which, among all 1D Decoder CNN variants, produces the best results for larger spatial scales, closely followed by CNN1D-$W_{map}$, which also presents the best overall performance for smaller scales. On the other hand, CNN1D is poorly suited to handle incomplete datasets involving large gaps, as can be concluded by its subpar performance with respect to CNN1D-$W_{map}$ and CNN1D-TL. Similarly to previous results, none of the 1D variants are capable of outperforming TFIT, which does indeed present a better contamination source removal performance for large spatial scales. CNN2D-TL, however, outperforms TFIT for larger spatial scales, at the expense of a slightly worst contamination source removal performance, with respect to CD, for smaller spatial scales. 
\subsubsection{Transfer learning for phase shift correction}
\label{sec:phshift}
\begin{figure}
	\centering
	\includegraphics[width=1\columnwidth]{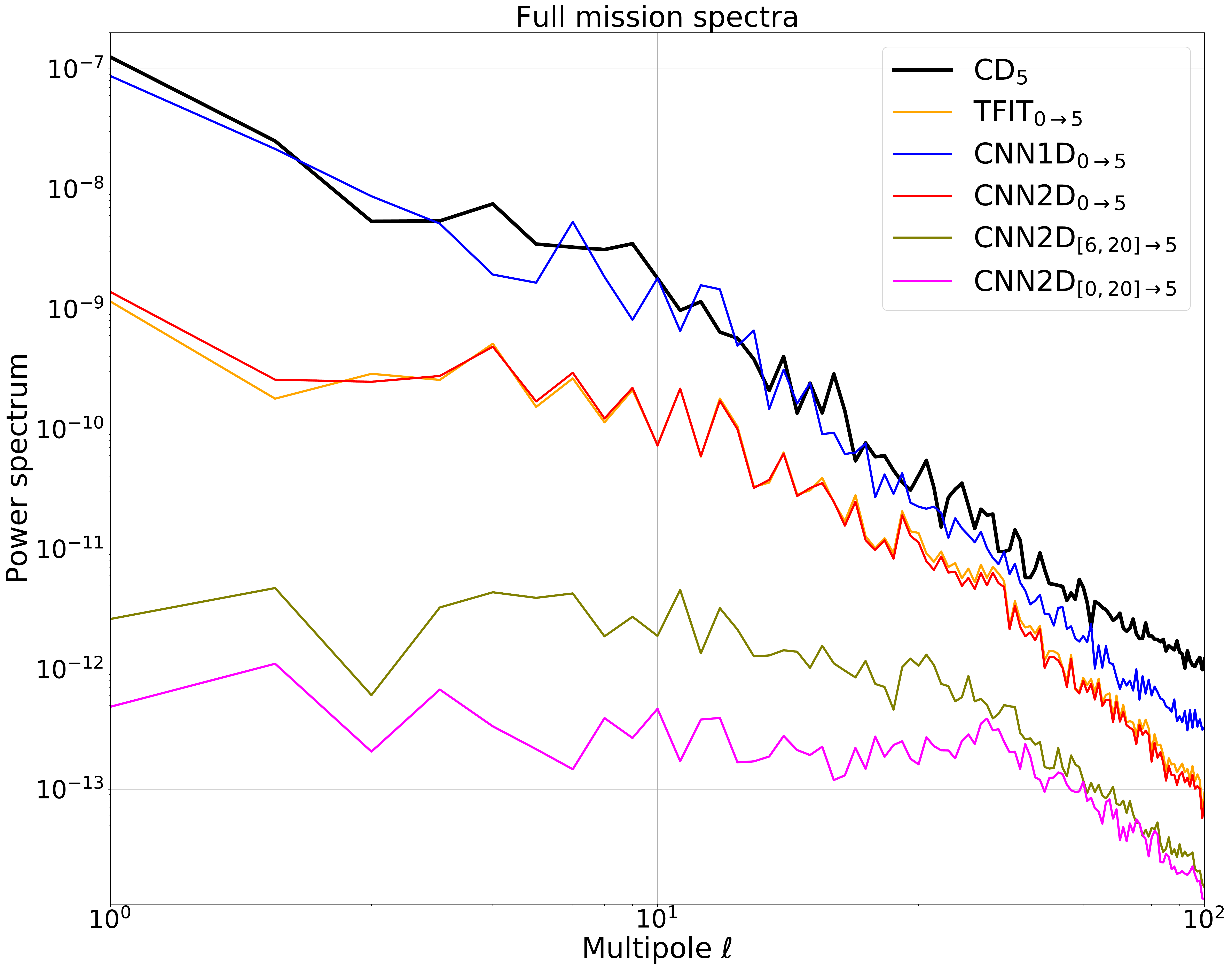}
	\caption{Power spectra of full mission maps of $5\degree$ phase-shifted detector $545_1$ FSL simulations after contamination source removal using 1000 iterations of a classic destriping approach (CD$_5$), a direct fit of non-shifted detector $545_1$ FSL simulations onto $5\degree$ phase-shifted detector $545_1$ FSL simulations as a template ( TFIT$_{0\rightarrow5}$), the 1D Decoder CNN trained on non-shifted detector $545_1$ FSL simulation data and applied to $5\degree$ phase-shifted detector $545_1$ FSL simulation data ( CNN1D$_{0\rightarrow5}$), the 2D Decoder CNN trained non-shifted on detector $545_1$ FSL simulation data and applied to $5\degree$ phase-shifted detector $545_1$ FSL simulation data ( CNN2D$_{0\rightarrow5}$), the 2D Decoder CNN trained on a catalog built from detector $545_1$ FSL simulation data shifted by $[6\degree,8\degree,\ldots,20\degree]$ and applied to $5\degree$ phase-shifted detector $545_1$ FSL Simulation data (CNN2D$_{[6,20]\rightarrow5}$), and the 2D Decoder CNN trained on a catalog built from detector $545_1$ FSL simulation data shifted by $[0\degree,2\degree,\ldots,20\degree]$ and applied to $5\degree$ phase-shifted detector $545_1$ FSL simulation data (CNN2D$_{[0,20]\rightarrow5}$).}
	\label{fig:fullhalf_tl}
\end{figure}

To further illustrate the relevance of transfer learning strategies to improve the characterization of large-scale systematic effects, we consider a case study involving 545 GHz FSL simulations with additional phase shift values. The primary objective is to evaluate the ability of the proposed approach to extract knowledge from an incomplete or inconsistent dataset that, nonetheless, contains relevant information that may be exploited to learn a suitable low-dimensional representation of the signals of interest. As previously explained, considering phase-shifted data at different phase shift values allows us to emulate both partially similar detectors as well as inaccurate FSL templates. As such, a phase-shifted version of the original FSL simulation is exploited to learn the Decoder CNN weights and biases, which are then subsequently applied to the mapmaking and contamination source removal of the original FSL simulation. Moreover, we also explore the possibility of combining multiple phase-shifted datasets as a means to construct an enriched dataset that better represents the relevant information to be learnt by the CNN. To this end, the weights and biases computed from the phase-shifted FSL are fixed, and only the low-dimensional inputs are retrained on the original (non-shifted) FSL observations.\\
We evaluate performance by presenting and comparing results obtained for the following approaches:
\begin{itemize}
	\item A classic destriping of $5\degree$ phase-shifted detector $545_1$ FSL simulation data (referred to as CD$_5$ hereafter),
	\item A direct fit of a FSL template computed from non-shifted detector $545_1$ FSL simulation data onto $5\degree$ phase-shifted detector $545_1$ FSL simulation data (referred to as TFIT$_{0\rightarrow5}$ hereafter),
	\item The Decoder CNN in its original 1D version trained on non-shifted detector $545_1$ FSL simulation data and applied to $5\degree$ phase-shifted detector $545_1$ FSL simulation data by retraining inputs only (referred to as CNN1D$_{0\rightarrow5}$ hereafter),
	\item The Decoder CNN in its 2D version trained on non-shifted detector $545_1$ FSL simulation data and applied to $5\degree$ phase-shifted detector $545_1$ FSL simulation data by retraining inputs only (referred to as CNN2D$_{0\rightarrow5}$ hereafter),
	\item The Decoder CNN in its 2D version trained on a catalog built from detector $545_1$ FSL simulation data shifted by $[6\degree,8\degree,\ldots,18\degree,20\degree]$ and applied to $5\degree$ phase-shifted detector $545_1$ FSL Simulation data by retraining inputs only (referred to as CNN1D$_{[6,20]\rightarrow5}$ hereafter),
	\item The Decoder CNN in its 2D version trained on a catalog built from detector $545_1$ FSL simulation data shifted by $[0\degree,2\degree,\ldots,18\degree,20\degree]$ and applied to $5\degree$ phase-shifted detector $545_1$ FSL Simulation data by retraining inputs only (referred to as CNN1D$_{[0,20]\rightarrow5}$ hereafter).
\end{itemize}
Given that we are interested in evaluating the potential of the transfer learning based 2D Decoder CNN to accurately learn the shape of FSL pickups, all considered networks rely on a single input for the low-dimensional representation of the signals of interest. The principle behind such an architecture is that the CNN weights and biases will capture the overall shape of FSL pickups, which implies that the free low-dimensional input should capture the phase shift between the different datasets considered. As such, the 2D architecture consists of an initial fully connected layer that projects a single input onto $K=32$ channels to produce a tensor of size $[1,4^{2},2 \cdot 4^{1},32]$, followed by $M-1$ 2D deconvolutional layers 
to dilate these $K$ channels and produce tensors of sizes $[16,8,K],[64,32,K],\ldots,[4^{m+1},2\cdot4^{m},K],\ldots,[4^{M+1},2\cdot4^{M},K]$, respectively. A circular deconvolutional layer 
then combines the existing $K$ channels to produce a tensor of size $[4^{M+2},2\cdot4^{M+1}]$. For training, time ordered data is binned into $4^{M+2}\times2\cdot4^{M+1}$ bins in ring and phase space, respectively, to match the network output.\\
For the present case study, we consider $M=3$, so that the proposed network outputs relies on a $1024\times512$ binning of time ordered data in ring and phase space. We present similar results as those introduced in previous sections, i.e., 
full mission power spectra in Fig. 
\ref{fig:fullhalf_tl}.\\
%
From Fig. \ref{fig:fullhalf_tl}, we conclude that CNN1D$_{0\rightarrow5}$ is only able to marginally improve contamination source removal performance (with respect to CD$_5$) for smaller spatial scales. This is expected, as the 1D variant of the Decoder CNN processes each ring independently and thus has a limited potential to model two-dimensional information, which appears as essential to accurately capture and model the phase difference to be transferred between the datasets involved. CNN2D$_{0\rightarrow5}$ performs similarly to TFIT$_{0\rightarrow5}$, and both approaches provide a significant contamination source removal performance improvement at all spatial scales. Such a result is explained by the fact that, since the CNN was trained on non-shifted data, it is unable to model phase shifts, as this is phenomena is not accurately represented in the training dataset. Indeed, contamination source removal performance is considerably increased when CNN2D$_{[6,20]\rightarrow5}$ is considered, which further supports the fact that the inclusion of phase-shifted data is necessary to ensure that the trained CNN learns to accurately represent phase shifts. Contamination source removal performance, particularly for larger spatial scales, is further improved with CNN2D$_{[0,20]\rightarrow5}$, when additional phase-shifted data (between $0\degree$ and $4\degree$) is considered. This is to be expected, as deep neural networks perform well for interpolation, but lack the necessary information to have similar performance for extrapolation. Adding phase-shifted data for smaller phase shift values means that the $5\degree$ phase shift of the target dataset is now inside the phase shift training range, and the trained network is better capable of modeling such phase shift.
\subsection{Application to 857 GHz data}
\label{sec:results_857}
\begin{figure*}
	\centering
	\includegraphics[width=1\textwidth]{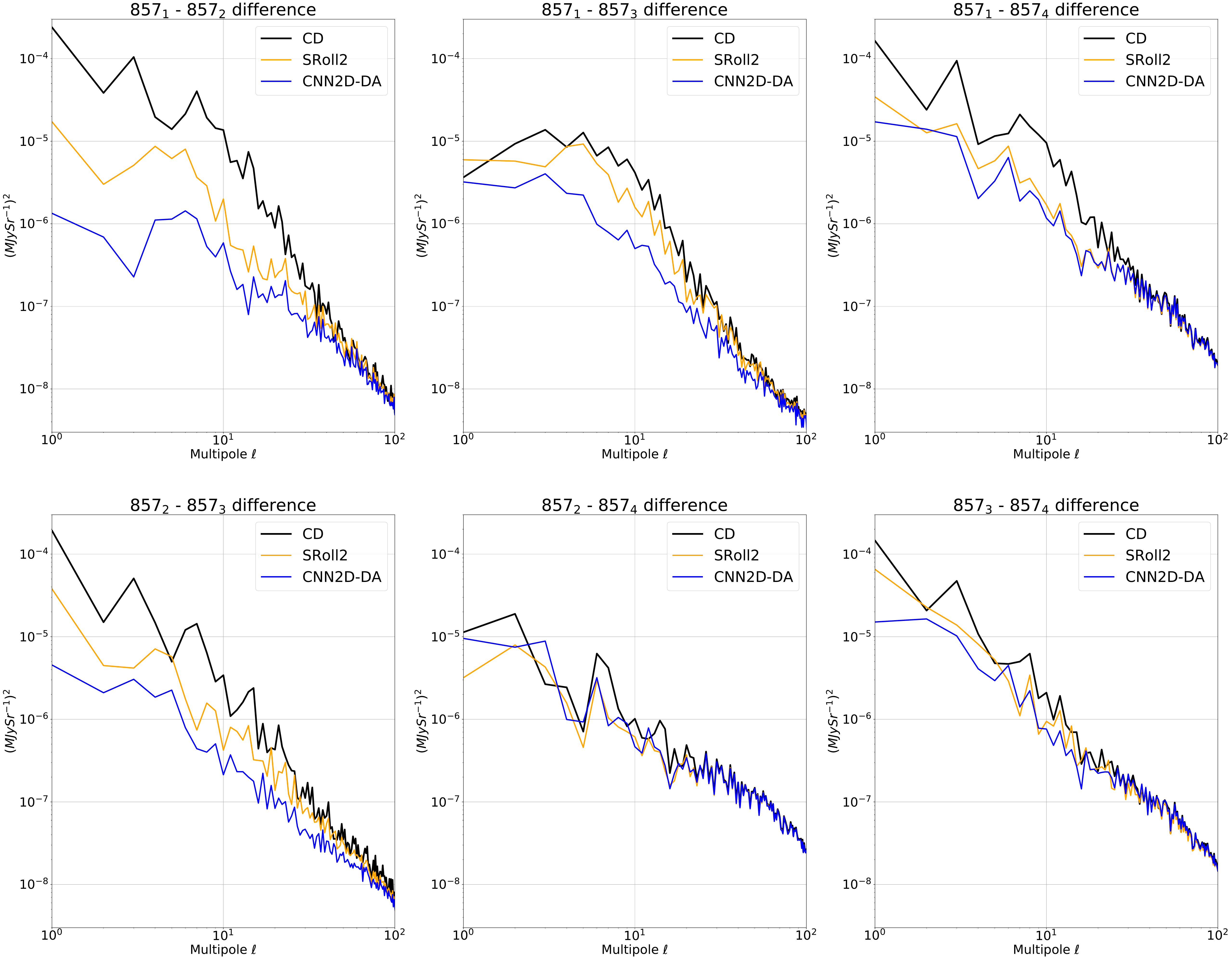}
	\caption{Power spectra for detector difference maps of Planck-HFI 857 GHz real data. For all detector pairs, power spectra are computed from detector difference maps for three distinct cases: the mapmaking of Planck-HFI 857 GHz real data using a classic destriping approach (CD), the mapmaking of Planck-HFI 857 GHz real data using SRoll2 to produce a direct fit of a synthetic FSL simulation as a template (SRoll2), and the mapmaking of Planck-HFI 857 GHz real data using 1000 iterations of the 2D Decoder CNN exploiting data augmentation to include inter-detector residuals in the learning dataset (CNN2D-DA).}	
	\label{fig:fullhalf_857}
\end{figure*}

\begin{figure*}
	\centering
	\includegraphics[width=1\textwidth]{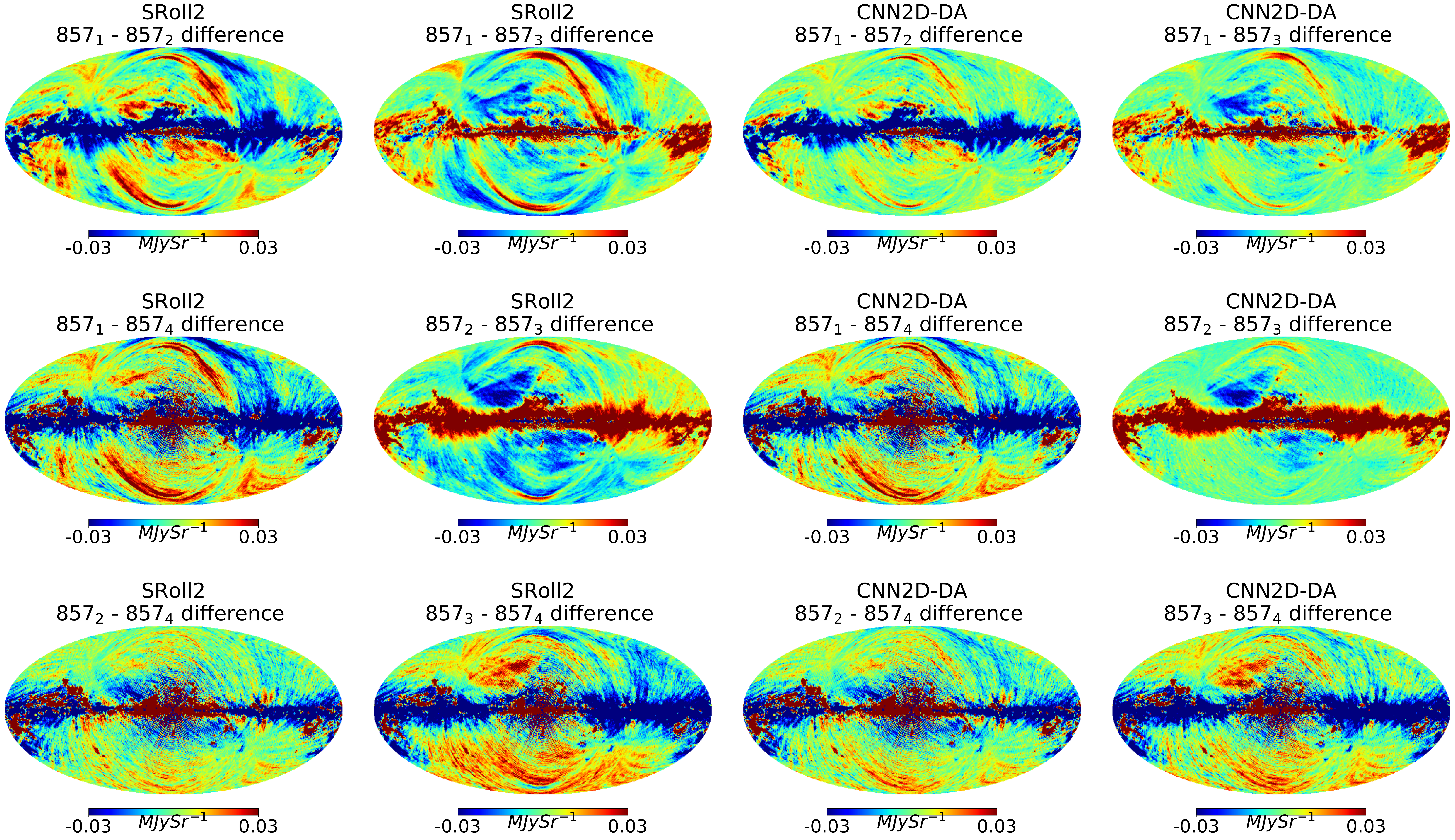}
	\caption{Detector difference maps of Planck-HFI 857 GHz real data. For all detector pairs, detector difference maps are computed for three distinct cases: the mapmaking of Planck-HFI 857 GHz real data using a classic destriping approach (not shown), the mapmaking of Planck-HFI 857 GHz real data using SRoll2 to produce a direct fit of a synthetic FSL simulation as a template (SRoll2, two leftmost columns), and the mapmaking of Planck-HFI 857 GHz real data using 1000 iterations of the 2D Decoder CNN exploiting data augmentation to include inter-detector residuals in the learning dataset (CNN2D-DA, two rightmost columns).}
	\label{fig:maps_857}
\end{figure*}

Following the validation of the proposed methodology on Planck-HFI 545 GHz FSL synthetic data, we evaluate its performance on real Planck-HFI 857 GHz observations. As previously explained, the Planck-HFI 857 GHz channel provides an ideal setting for evaluating the ability of the proposed approach to model and remove large-scale systematic effects, and FSL pickup in particular, given that the detector difference between the four 857 GHz detectors will mostly depict large-scale systematic effects, and predominantly the FSL pickup \citep{planck2016-l03p}. The 857 GHz dataset consists of time ordered data from four independent detectors (named hereafter $857_d,d=1,\ldots,4$).\\
In this context, we exploit the 2D variant of the Decoder CNN using a single input for the low-dimensional representation of the signals of interest.The 2D architecture is then identical to the one used to explore the potential of transfer learning techniques to correct phase shifts in Sect. \ref{sec:phshift}, and consists of an initial fully connected layer that projects a single input onto $K=32$ channels to produce a tensor of size $[4^{2},2 \cdot 4^{1},K]$, followed by $M-1$ 2D deconvolutional layers 
to dilate these $K$ channels to produce tensors of sizes $[64,32,K],\ldots,[1,4^{m+1},2\cdot4^{m},\ldots,[1,4^{M+1},2\cdot4^{M},K]$, respectively. A circular deconvolutional layer 
combines the existing $K$ channels to produce a tensor of size $[4^{M+2},2\cdot4^{M+1}]$. For training, time ordered data is binned into $4^{M+2}\times2\cdot4^{M+1}$ bins in ring and phase space, respectively, to match the network output.\\
We also explore the potential of data augmentation to integrate expert knowledge into the training of the Decoder CNN and thus provide enhanced modeling capabilities for the FSL pickup. To this end, the training dataset is enhanced by integrating information from all four detectors into the contamination source removal procedure of each individual detector. Specifically, for each detector, the training dataset in enriched by integrating the residue of detectors $857_1$, $857_3$ and $857_4$ with respect to detector $857_2$. detector $857_1$ is chosen as the common base for all residues considered simply because its position within the detector array effectively reduces its FSL pickup. 
The computation of these residues is performed after the data is binned in ring and phase spaces. We consider $M=3$, so that time ordered data is initially binned into 1024 bins in ring space and 512 bins in phase space. Once datasets for the four detectors have been binned, each detector dataset is enriched by adding the residue, i.e., the difference, between detector $857_2$ binned data and binned data from the three remaining detectors. These residues are then subjected to a thresholding procedure, such that all data whose absolute value is below a user-set threshold is set to 0. The idea behind this procedure is that the considered residues will not only contain relevant FSL pickups that can be used for training the Decoder CNN, but also other noise signals that should not be taken into account and that should, ideally, be filtered by the thresholding operation. As such, a coarse value for the threshold is set empirically by taking into account the noise levels within the considered dataset, and then fine-tuned by performing multiple simulations at different threshold values. Given that the threshold is user-set, this procedure can be seen as the integration of expert knowledge into the otherwise non-supervised procedure of network training. The final approach could therefore be qualified as a weakly supervised network training method. The proposed augmented datasets are used to train the Decoder CNN weights and biases (independently for each detector), with network inputs then being retrained directly on the original non-augmented detector datasets.\\
As previously explained, Planck-HFI 857 GHz detector difference maps are dominated by the FSL pickup signal, which makes them an ideal gauge for the capacity of the proposed approach to remove the FSL pickup from the final maps. Taking this into account, we illustrate our results by presenting the power spectra of Planck-HFI 857 GHz detector difference maps in Fig. \ref{fig:fullhalf_857}, and the Planck-HFI 857 GHz detector difference maps themselves in Fig. \ref{fig:maps_857}. For visualization and comparison purposes, all detector difference maps are normalized to a common baseline amplitude level, and any existing CO difference map signatures are removed using the same template fit procedure used by SRoll2 to produce the 2018 release of the Planck-HFI sky maps. We compare results for three different cases, namely:
\begin{itemize}
	\item The mapmaking of Planck-HFI 857 GHz real data using a classic destriping approach (referred to as CD hereafter),
	\item The mapmaking of Planck-HFI 857 GHz real data using SRoll2 \citep{delouis2019p} to produce a direct fit of a synthetic FSL simulation as a template (referred to as SRoll2 hereafter),
	\item The mapmaking of Planck-HFI 857 GHz real data using 1000 iterations of the 2D Decoder CNN exploiting data augmentation to include inter-detector residuals in the learning dataset (referred to as CNN2D-DA hereafter).
\end{itemize}
From power spectra depicted in Fig. \ref{fig:fullhalf_857}, we can conclude that the inter-detector data augmentation strategy, coupled with the introduction of expert knowledge via the thresholding of binned data residues, allows for a considerable improvement in contamination source removal performance for all spatial scales and for most detector pairs, with a considerable gain for larger spatial scales. As such, as far as large-scale contamination source removal is concerned, CNN2D-DA seems to outperform SRoll2 for most detector pairs. Indeed large-scale contamination source removal performance is only marginally degraded for a single detector pair ($857_2-857_4$) and only for larger spatial scales (around $\ell<20$). For the remaining detector pairs, we report considerable gains, typically up to one order of magnitude, in terms of contamination source removal performance for large spatial scales. This demonstrates the relevance of the proposed methodology to correctly capture and remove the FSL pickup signal during the data inversion. 
These conclusions are further supported by the detector difference maps presented in Fig. \ref{fig:maps_857}, where one can observe a considerable improvement in contamination source removal performance for most detector pairs, with respect to a SRoll2, for CNN2D-DA. Interestingly, a particularly strong large-scale signal can be observed to the north of the galactic plane, near the galactic origin. Given that we are working with Planck-HFI 857 GHz real data, we hypothesize this signal to be caused by other contamination sources, which explains CNN2D-DA inability to completely remove it, as it has been extensively adapted, in the presented application, to deal specifically with FSL pickups.
\section{Discussion}
\label{sec:discussion}
\subsection{Data-driven modeling of systematic effects}
\subsubsection{Map constraint}
As explained in Sect. \ref{sec:map_constraint}, the Decoder CNN may introduce an erroneous large-scale signal to its reconstructed output. Indeed, since the Decoder CNN is trained on signal co-occurrences only, cost function (\ref{eq:loss}) may be artificially decreased by adding an adequately chosen large-scale offset, whereas the introduction of this offset does not necessarily relate to the contamination source removal of the final map. According to our results, such large-scale signature may appear in the form of a large-scale offset, or even higher order moments, such as a large-scale spatial dipole. In particular, this was observed for results presented in Sect. \ref{sec:tl_results}, specifically for the case considering one every ten rings, i.e., for partial datasets involving large gaps. This is expected, given that in such cases the lack of observations between the ecliptic poles is exacerbated, thus further strengthening this effect. As can be observed in our results, the introduction of a map constraint (Eq. (\ref{eq:constraint_map})) helps limit the introduction of a large-scale offset, given that it improves the network conditioning in difficult cases, such as those considering partial, gap-filled or irregularly sampled datasets. This demonstrates both the flexibility of the proposed framework to be adapted to the dataset and/or problem to be treated by incorporating appropriate additional terms to the custom cost function (\ref{eq:loss}), as well as its capability to adequately handle partial, gap-filled datasets.
\subsubsection{Transfer learning}
As observed, the exploitation of transfer learning techniques allows for the characterization of the ``shape" of the systematic effects or foregrounds we are trying to separate from our signal of interest. This is achieved by constraining the smaller dimensional subspace onto which the consider signals are projected. The ``shape" of systematic effects and foregrounds is indeed encoded into a projection operator, which is parameterized by the Decoder CNN, by minimizing the loss function on the training dataset. The trained Decoder CNN is then applied to a second dataset by retraining the inputs only. As previously stated, this can be seen as a way of identifying and learning the common knowledge between the different datasets (i.e., the projection) and transferring such knowledge between different datasets. Such an approach is particularly relevant for applications where similar foregrounds or systematic effects exist between different datasets, as is the case for the FSL pickup.
Indeed, in the presented application, the Decoder CNN training stage seems to learn general characteristics of the FSL pickup signal, such as its large-scale signature, which is then transferred to the second dataset (by retraining inputs) in order to improve contamination source removal performance. From a mathematical point of view, retraining the inputs can be thought of as finding the representation in the projection subspace that best approximates the second dataset. This amounts to finding the best fitting FSL pickup signal approximation, under the constraint that the characteristics of this approximation were previously learned on the first dataset and encoded in the Decoder CNN weights and biases.
\subsection{Neural network based removal of large-scale systematic effects}
\subsubsection{Data augmentation}
Results obtained for Planck-HFI 857 GHz real data illustrate how data augmentation techniques, coupled with expert knowledge integration, can improve contamination source removal performance. Indeed, introducing, for each 857 GHz detector, inter-detector residuals with respect to detector $857_2$, comes to exploiting data augmentation to transfer relevant information between datasets. As such, this procedure closely relates to the idea of transfer learning, since both seek to exploit information shared between datasets to improve contamination source removal performance. Moreover, the inclusion of a user-set threshold for inter-detector binned data residues allows us to integrate expert knowledge into an otherwise completely unsupervised learning scheme. This is particularly relevant for the processing of data containing both well-known and badly modeled signals, as is the case for 
the systematic effects and foregrounds present in Planck-HFI observations.
%
%
\section{Conclusions}
\label{sec:conclusion}
In the present work, we propose a neural network based data inversion approach to reduce structured contamination sources, with a particular focus on the mapmaking for Planck-HFI data and the removal of large-scale systematic effects within the produced sky maps. The proposed approach relies on an generative decoder convolutional neural network to project the signals of interest onto a learned low-dimensional subspace simultaneously with the data inversion, so that the low-dimensional subspace is optimized with respect to the contamination source removal and mapmaking objectives. This optimization is achieved by means of a loss function that take such objectives into account during the network training stage. The exploitation of such a custom loss function also allows for the introduction of physics-based constrains to further improve contamination source removal performance. The low-dimensional subspace learning is possible thanks to an input-training scheme, which also allows for the processing of incomplete and/or gap-filled datasets. We propose multiple variants of the proposed approach, a two-dimensional version capable of taking time dependencies into account, as well as variants exploiting transfer learning, data augmentation and the introduction of expert knowledge to further improve reconstruction performance. Importantly, the proposed method is capable of exploiting spatiotemporal scale couplings within contamination sources to learn, simultaneously with the data inversion, a low-dimensional representation that facilitates the removal of these contamination sources. Whereas this is illustrated here for an example considering Planck-HFI data, the method provides a general framework for structured contamination source removal, and may be used to tackle similar problems in other scientific contexts. Indeed, the proposed approach can potentially be applied to any data inversion problem dealing with contamination sources, provided that these sources are sufficiently structured to allow for the determination of a suitable low-dimensional subspace, optimized to facilitate the data inversion.\\
We validate the proposed approach on synthetic 545 GHz Planck-HFI data comprising simulated FSL pickups. This validation on synthetic datasets demonstrates the relevance of the two-dimensional variant of the proposed approach to better remove FSL pickup signals simultaneously with the data inversion, with respect to both a classic destriping approach as well as the direct fit of simulated FSL pickups as a template, particularly for partial, gap-filled observation datasets (comprising a subsampling of one every ten rings). Moreover, the relevance of the two-dimensional variant to efficiently exploit transfer learning approaches to model and capture phase-shifts in observations is also demonstrated during the validation on synthetic simulated data.\\
Following validation, we further explore the proposed approach by applying it to the contamination source removal and mapmaking of real 857 GHz Planck-HFI observations. We exploit the two-dimensional variant of the proposed method, alongside with data augmentation, to demonstrate the relevance of the proposed framework to outperform both a classic destriping approach as well as a direct fit of FSL pickup as a template for the removal of large-scale systematic effects in real data. In particular, the case study clearly depicts how inter-detector data augmentation and the integration of expert knowledge, by means of a user-set threshold for noise removal in the augmented dataset, allows for a considerable gain in terms of FSL pickup removal, thus improving mapmaking and contamination source removal performance.\\
Generally speaking, the present work underlines the relevance of data-driven neural network based approaches to improve on current contamination source removal and mapmaking approaches and go beyond their limitations by providing enhanced capabilities for the separation and removal of structured, non-Gaussian information, such as systematic effects and foregrounds, which should allow for the creation of more accurate CMB maps and thus improve current parameter likelihood estimates in order to better constraint and/or validate cosmological models.\\
Importantly, this work builds on previously developed methods for the separation and removal of structured contamination sources, and particularly on the SRoll2 algorithm \citep{delouis2019p}. As such, the methods developed in this work are to be integrated in a new version of the SRoll algorithm (SRoll3), and we describe here SRoll3 857 GHz detector maps that will be released to the community.
\subsection{Future work}
The possible research avenues stemming from the proposed approach include a wide arrange of both theoretical and practical issues. Whereas, in this work, we illustrated the relevance of the proposed approach for the modeling and removal of systematic effects, 
we underline the suitability of the proposed methodology for the modeling and removal of any structured signal, including modeling errors, observation errors, and foregrounds, among others. This implies that the proposed framework can be applied to a wide range of similar problems in multiple scientific domains, ranging from the mapmaking and contamination source removal of Planck-HFI data to the removal of structured noise sources in new generation ocean remote sensing satellite missions, or even the processing of ground-borne and balloon-borne sky observations. Furthermore, one may also consider, for example, exploiting the proposed Decoder CNN to apply transfer learning techniques to the component separation problem in Planck data. 
In this regard, a multi-channel Decoder CNN could be exploited to separate different components, with different channels representing different sources. In this context, transfer learning techniques could be used on specific channels to better capture the source considered, similarly to the approach illustrated above for the FSL foreground. The modeling and correction of Analog-to-Digital Converter (ADC) non-linearities \citep{planck2014-a08} also appears as a current issue that could greatly benefit from the proposed transfer learning based formulation. Indeed, we expect that exploiting transfer learning should allow us to better understand and model the ADC non-linearities that exist within the Planck-HFI data by exploiting simulated and/or real data to learn a low-dimensional representation where such non-linearities may become easier to correct. Finally, the processing of ground-based cosmological observations may also be considered as a potential application of the proposed approach, particularly with respect to the removal of atmospheric turbulence related noise, given its slow temporal variation.
\begin{acknowledgements}
	This work is part of the Bware project supported by CNES, and part of the Deepsee project supported by the Programme National de Télédétection Spatiale of the CNRS Institut des Sciences de l'Univers (http://www.insu.cnrs.fr), grant n° PNTS-2020-08. The authors acknowledge the heritage of the Planck-HFI consortium regarding data, software, knowledge. The program was granted access to the HPC resources of CINES (http://www.cines.fr) under the allocation 2020-A0080411364 made by GENCI (http://www.genci.fr). MLR acknowledges the financial support of the ``Chaire de Cosmologie'' of the Fondation de l'Université Paris-Saclay (https://www.fondation.universite-paris-saclay.fr).
\end{acknowledgements}
\bibliography{SRoll3.bib}
\typeout{get arXiv to do 4 passes: Label(s) may have changed. Rerun}
\end{document}